\documentclass{article} 
\usepackage{epsfig} 
\usepackage{cite}  
\def\H{\hbox{H}}

\def\d{\hbox{d}} 
\def\dd{\partial} 
\def\f{\hbox{f}} 
\def\FORM{{\tt FORM}} 
\def\g{\hbox{g}} 
 
\def\Li{\hbox{Li}} 
\def\ln{\hbox{ln}}

\def\G{\hbox{G}} 
\def\Gp{\hbox{G}_+} 
\def\Gm{\hbox{G}_-} 
\def\ie{{\it i.e.}}

\begin{document} 
\unitlength1cm 
\begin{titlepage} 
\vspace*{-1cm} 
\begin{flushright} 
CERN-TH/2001-326\\ 
hep-ph/0111255\\ 
November 2001 
\end{flushright} 
\vskip 3.5cm 

\begin{center} 
\boldmath 
{\Large\bf Numerical Evaluation of\\[2mm] Two-Dimensional Harmonic 
Polylogarithms }\unboldmath 
\vskip 1.cm 
{\large  T.~Gehrmann}$^a$ and {\large E.~Remiddi}$^b$ 
\vskip .7cm 
{\it $^a$ Theory Division, CERN, CH-1211 Geneva 23, Switzerland} 
\vskip .4cm 
{\it $^b$ Dipartimento di Fisica, 
    Universit\`{a} di Bologna and INFN, Sezione di 
    Bologna,  I-40126 Bologna, Italy} 
\end{center} 
\vskip 2.6cm 

\begin{abstract} 
The two-dimensional harmonic polylogarithms $\G(\vec{a}(z);y)$, a 
generalization of the harmonic polylogarithms, themselves a generalization 
of Nielsen's polylogarithms, appear in analytic calculations of 
multi-loop radiative corrections in quantum field theory. 
We present an algorithm for the numerical evaluation of two-dimensional 
harmonic polylogarithms, with the two arguments $y,z$ varying in the 
triangle $0\le y \le 1$, $\ 0\le z \le 1$, $\ 0\le (y+z) \le 1$. 
This algorithm is implemented into a {\tt FORTRAN} subroutine {\tt tdhpl} 
to compute two-dimensional harmonic polylogarithms up to weight 4.
\end{abstract} 
\vfill 
\end{titlepage} 
\newpage 

{\bf\large PROGRAM SUMMARY}
\vspace{4mm}
\begin{sloppypar}
\noindent   {\em Title of program\/}: {\tt tdhpl} \\[2mm]
   {\em Version\/}: 1.0 \\[2mm]
   {\em Release\/}: 1  \\[2mm]
   {\em Catalogue number\/}: \\[2mm]
   {\em Program obtained from\/}:
   {\tt Thomas.Gehrmann@cern.ch, Ettore.Remiddi@bo.infn.it} \\[2mm]
   {\em E-mail\/}: {\tt Thomas.Gehrmann@cern.ch, Ettore.Remiddi@bo.infn.it} \\[2mm]
   {\em Licensing provisions\/}: no \\[2mm]
   {\em Computers\/}: all \\[2mm]
   {\em Operating system\/}: all \\[2mm]
   {\em Program language\/}: {\tt FORTRAN77     } \\[2mm]
   {\em Memory required to execute\/}: Size: 1144k \\[2mm]
   {\em No.\ of bits per word\/}: up to 32 \\[2mm]
   {\em No.\ of lines in distributed program\/}: 34589 \\[2mm]
   {\em Other programs called\/}: {\tt hplog} (from the same authors, 
Comput.\ Phys.\ Commun.\ {\bf 141} (2001) 296) \\[2mm]
   {\em External files needed\/}: none \\[2mm]
   {\em Nature of the physical problem\/}: 
        numerical evaluation of the two-dimensional 
harmonic polylogarithms up to weight 4 
        for real arguments restricted to $0\leq y \leq 1$, 
        $\ 0\leq z \leq (1-y)$.  These functions are emerging in 
        Feynman graph integrals involving three mass scales. \\[2mm]
   {\em Method of solution\/}: 
        for small values of the argument: series representation 
        with expansion coefficients depending on the second argument; 
        other values of the argument: transformation formulae. \\[2mm] 
   {\em Restrictions on complexity of the problem\/}: limited  
        to 2dHPLs of up to weight 4, 
        the algorithms used here can be extended to 
        higher weights without further modification. \\[2mm]
   {\em Typical running time\/}:
        on average 1.8 ms (both arguments varied)/0.5 ms (only one argument 
        varied) for the evaluation of all 
        two-dimensional harmonic polylogarithms up 
        to weight 4 on a Pentium III/600 MHz Linux PC. 
\end{sloppypar}

\newpage

{\bf\large LONG WRITE-UP}

\renewcommand{\theequation}{\mbox{\arabic{section}.\arabic{equation}}} 

\section{Introduction} 
\label{sec:int} 
\setcounter{equation}{0}
The two-dimensional harmonic polylogarithms, a generalization of the 
harmonic polylogarithms $\H(\vec a,x)$, have been introduced in 
\cite{doublebox} 
for the analytic evaluation of a class of two-loop, off-mass-shell 
scattering Feynman graphs in massless QCD. 

Two-dimensional harmonic polylogarithms are obtained by the repeated 
indefinite integration of rational fractions, involving one further 
independent variable besides the integration variable. The first 
appearance of functions of this type in the mathematical literature was in a 
series of articles of E.E.~Kummer~\cite{kummer} in 1840. Allowing 
for arbitrarily many different variables to appear in the rational fractions, 
one obtains a class of functions introduced by Poincar\'{e}, which are 
called `hyperlogarithms'. These have been studied in great 
detail in the works of J.A.\ Lappo-Danilevsky~\cite{Lappo}. 
Hyperlogarithms and their generalization `multiple polylogarithms' are 
receiving renewed attention in the mathematical literature (see for 
instance the review of A.B.\ Goncharov~\cite{gonch}). 

It has been known for a long time that 
the analytic evaluation of integrals in 
perturbative quantum field theory gives rise to 
the Euler dilogarithm ${\rm Li}_2(x)$ and its generalizations, 
Nielsen's polylogarithms~\cite{Nielsen}. A reliable and 
widely used numerical representation of these functions 
({\tt GPLOG})~\cite{bit}
has been available for already thirty years. Going to higher 
orders in perturbation theory, it was recently realized that 
Nielsen's polylogarithms are insufficient to evaluate all integrals 
appearing in Feynman graphs at two loops and beyond. This limitation 
can only be overcome by the introduction of further generalizations 
of Nielsen's polylogarithms. This 
generalization is made by the harmonic polylogarithms (HPLs)~\cite{hpl},
appearing in loop integrals involving two mass scales, and 
two-dimensional harmonic polylogarithms (2dHPLs)~\cite{doublebox}, 
appearing in loop integrals involving three mass scales.  
These functions are already now playing a 
central role in the analytic evaluation of Feynman graph 
integrals~\cite{doublebox,moch}. HPLs appear moreover as inverse 
Mellin transformations of harmonic sums, 
which were investigated and implemented numerically in~\cite{blumlein},
while 
2dHPLs and multiple polylogarithms also 
appear for example if generalized hypergeometric functions 
are expanded in their indices around integer values~\cite{muw}. \par 
A subroutine ({\tt hplog}) for the 
numerical evaluation of HPLs for arbitrary, real values of the argument 
was presented in~\cite{1dhpl}.  

In this paper, as a continuation of \cite{1dhpl}, 
 we briefly review the analytical properties of the 2dHPLs and then show 
how those properties can be used for writing a {\tt FORTRAN} 
code that evaluates the 2dHPLs  up to weight 4 (see Section~\ref{sec:def} 
below for the definition of the weight; 4 is the maximum weight 
required in the calculations of \cite{doublebox}) with absolute 
precision better than $3\times 10^{-15}$ (\ie \ standard double precision) 
with a few dozens of elementary arithmetic operations per function.
Given the large number (256) of 2dHPLs  of weight 4, the many algebraic 
relations among them, and the fact that any application is likely to involve 
a large number of them at the same time (see for instance the results of 
\cite{doublebox}) our {\tt FORTRAN} routine evaluates the whole set of 
2dHPLs up to the required weight -- as in {\tt hplog}~\cite{1dhpl},
but at variance with {\tt GPLOG}~\cite{bit}, which evaluates separately 
(and up to weight 5) the various Nielsen polylogarithms. While {\tt hplog}
evaluated the HPLs for arbitrary real value of the argument, yielding 
complex results, we 
restrict the arguments of the 2dHPLs to the region $0\leq y\leq 1-z$,
$0\leq z\leq 1$, where the 2dHPLs are real. This region is the only relevant 
region for applications in quantum field theory~\cite{doublebox}. 

The plan of the paper is as follows. 
Section~\ref{sec:def} recalls the definitions of the HPLs. Their
algebraic properties are discussed in 
Section~\ref{sec:alg}, where we show how to use these properties 
for separating the functions into reducible and irreducible ones.
In Section~\ref{sec:values}, we discuss the behaviour of 2dHPLs for 
special values of the argument.
Relations between 2dHPLs  for different ranges of the arguments 
are derived in Section~\ref{sec:ids}. 
Section~\ref{sec:anal} studies the analytic properties that allow the
performing of converging power series expansions and the acceleration
of  their 
convergence. 
Section~\ref{sec:numer} explains  how the properties recalled above 
are used to implement the 2dHPLs into the {\tt FORTRAN} subroutine 
{\tt tdhpl}, and Section~\ref{sec:checks} 
how the correct implementation is checked. 
Finally, we describe the usage of the subroutine {\tt tdhpl} in 
Section~\ref{sec:code} and provide a few numerical examples in 
Section~\ref{sec:plots}. We enclose two appendices. 
Appendix~\ref{app:notation} compares 
the notations used for 2dHPLs in different previous publications and 
Appendix~\ref{app:nielsen} provides relations between particular cases 
of the 2dHPLs and Nielsen's generalized polylogarithms. 

\section{Definitions} 
\label{sec:def} 
\setcounter{equation}{0} 
The 2dHPLs family which we 
consider here is obtained by the repeated integration, in the variable $y$, 
of rational factors chosen, in any order, from the set $1/y$, $1/(y-1)$, 
$1/(y+z-1)$, $1/(y+z)$, where $z$ is another independent variable (hence 
the `two-dimensional' in the name). It is clear that the set of rational 
factors might be further extended or modified; for the harmonic 
polylogarithms discussed in \cite{1dhpl} the set of rational factors was 
for instance $1/y$, $1/(y-1)$, $1/(y+1)$, involving only constants and no 
other variable besides $y$. \par 
More precisely and in full generality, let us define the rational factor 
as 
\begin{equation} 
  \g(a;y) = \frac{1}{y-a} \ , 
\label{eq:gya} 
\end{equation} 
where $a$ is the {\it index}, which can depend on $z$, $a=a(z)$; 
the rational factors which we consider for the 2dHPLs are then 
\begin{eqnarray} 
  \g(0;y) &=& \frac{1}{y} \ ,       \nonumber\\ 
  \g(1;y) &=& \frac{1}{y-1} \ ,     \nonumber\\ 
  \g(1-z;y) &=& \frac{1}{y+z-1} \ , \nonumber\\ 
  \g(-z;y) &=& \frac{1}{y+z} \ . 
\label{eq:gyalist} 
\end{eqnarray} 
With the above definitions the index takes one of the values 
$0, 1, (1-z) $ and $(-z)$. \par 
Correspondingly, the 2dHPLs at weight $w=1$ (\ie ~depending, besides 
the variable $y$, on a single further argument, or {\it index}) are 
defined to be 
\begin{eqnarray} 
 \G(0;y) &=& \ln\, y  \ ,                                \nonumber\\ 
 \G(1;y) &=& \ln\, (1-y) \ ,                            \nonumber\\ 
 \G(1-z;y) &=& \ln\left( 1 - \frac{y}{1-z} \right) \ , \nonumber\\ 
 \G(-z;y) &=& \ln\left( 1 + \frac{y}{z} \right) \ . 
\label{eq:w1list} 
\end{eqnarray} 
The 2dHPLs of weight $w$ larger than 1 depend on a set of $w$ indices, which 
can be grouped into a 
$w$-dimensional {\it vector} of indices $\vec{a}$. By writing the vector 
as $ \vec{a} = (a, \vec b)$, where $a$ is the leftmost component of 
$ \vec{a} $ and $\vec b $ stands for the vector of the remaining $(w-1)$ 
components, the 2dHPLs are then defined as follows: if all the $w$ 
components of $\vec a$ take the value 0, $\vec a$ is written as $\vec 0_w$ and 
\begin{equation} 
  \G(\vec{0}_w;y) = \frac{1}{w!} \ln^w{y} \ , 
\label{eq:defh0} 
\end{equation} 
while, if $\vec{a} \neq \vec{0}_w$, 
\begin{equation} 
  \G(\vec{a};y) = \int_0^y \d y' \ \g(a;y') \ \G(\vec{b};y') \ . 
\label{eq:defn0} 
\end{equation} 
In any case the derivatives can be written in the compact form 
\begin{equation} 
\frac{\dd }{\dd y} \G(\vec{a};y) = \g(a;y) \G(\vec{b};y) \ , 
\label{eq:derive} 
\end{equation} 
where, again, $a$ is the leftmost component of $ \vec a $ and 
$ \vec b $ stands for the remaining $(w-1)$ components. 

From (\ref{eq:defh0}) and (\ref{eq:defn0}), one arrives immediately at a 
multiple (or repeated) integral representation of the 2dHPL:
\begin{equation}
\G(\vec{m}_w;y) 
= \int_0^y \d t_1 \g(m_1;t_1) \int_0^{t_1} \d t_2 \g(m_2;t_2)\ldots
\int_0^{t_{w-1}} \d t_w \g(m_w;t_w)\; , 
\label{eq:zlimone}
\end{equation}
valid for $m_w\neq 0$, and
\begin{equation}
\G(\vec{m}_w;y) 
= \int_0^y \d t_1 \g(m_1;t_1) \int_0^{t_1} \d t_2 \g(m_2;t_2)\ldots
\int_0^{t_{v-1}} \d t_v \g(m_v;t_v) \G(\vec{0}_{w-v};t_v)\; , 
\label{eq:zlimtwo}
\end{equation}
valid for $\vec{m}_w = (\vec{m}_v,\vec{0}_{w-v})$ with 
$\vec{m}_v\neq \vec{0}_v$. 
\par 
The definition 
is essentially the same as for the harmonic polylogarithms of~\cite{1dhpl}, 
if allowance is made for the greater generality of the `indices', which can 
now depend on the second variable $z$. Let us, however, stress an important 
difference between the present definitions and the notation already used 
in \cite{doublebox}, where the rational factors were indicated 
by $\f(a,x)$ and the harmonic polylogarithms by $\H(\vec a,x)$; we have 
indeed 
\begin{eqnarray} 
  \f(1;x) &=& - \g(1;x) \ ,      \nonumber\\ 
  \f(1-z;x) &=& - \g(1-z;x) \ ,  \nonumber\\ 
  \f(z;x) &=& \g(-z;x) \ ,       \nonumber\\  
\label{eq:corr1} 
\end{eqnarray} 
while there is no change when $a=0$:
\begin{eqnarray} 
  \f(0;x) &=& \g(0;x) \ . 
\label{eq:corr2} 
\end{eqnarray} 
Also for $a=-1$ one would have $\f(-1;x) = \g(-1;x)$, but we 
will not consider this case here as it never appears together with 
the other values of the indices $(1-z), (-z)$ in \cite{doublebox}.
The same applies between the harmonic polylogarithms $\H$ previously 
introduced~\cite{hpl} 
and the 2dHPLs, as any $\H$-function of \cite{doublebox} goes 
into the corresponding $\G$-function, with the following correspondence 
rules: the indices $(1), (1-z)$ of $\H$ remain unchanged as indices of 
$\G$, but each occurrence of $(1), (1-z)$ gives a change of sign 
between $\H$ and $\G$; 
any index $(z)$ of $\H$ goes into an index $(-z)$ of $\G$ (which keeps the 
same sign as $\H$). One has for instance 
\begin{eqnarray} 
  \H(z,1-z;y) &=& - \G(-z,1-z;y) \ , \nonumber\\ 
  \H(0,z,1-z,1;y) &=& \G(0,-z,1-z,1;y) \ , 
\label{eq:correx} 
\end{eqnarray} 
and so on. 
The main advantage of the new notation is the (obvious) continuity in $z$ 
of the $\g$'s and the $\G$'s; one has for instance 
\begin{equation} 
    \lim_{z\to1} \g(1-z;y) = \g(0;y) \ , 
\label{eq:zcont} 
\end{equation} 
to be compared with 
\begin{equation} 
    \lim_{z\to1} \f(1-z;y) = - \f(0;y) \ , 
\label{eq:zdiscont} 
\end{equation} 
and the same applies to any index of a $\G$-function (when the limit 
exists). Note, however, that the positivity for positive value of the 
argument is lost -- so that, for instance, one has $\G(0,1;1) = - \pi^2/6$, 
to be compared with the more elegant relation $\H(0,1;1) = \pi^2/6$. 

The 2dHPLs can also be viewed as a special case of the hyperlogarithms, 
which are discussed frequently in the mathematical literature. 
We summarize the various available notational conventions in 
Appendix A, where we also provide appropriate translation rules. 

\section{The algebra and the reduction equations} 
\label{sec:alg} 
\setcounter{equation}{0}
Algebra and reduction equations of the 2dHPLs are the same as for the 
ordinary HPLs. They were discussed extensively in~\cite{hpl,1dhpl}, and 
identical formulae apply regardless of the actual values of the 
indices. In the following, we briefly summarize the results of 
\cite{hpl,1dhpl}, without providing explicit examples.

The product of two 2dHPLs  of a same argument $x$ 
and weights $p, q$ can be expressed as a combination of 2dHPLs  of that 
argument and weight $r=p+q$, according to the product identity 
\begin{eqnarray} 
 \G(\vec{p};x)\G(\vec{q};x) & = & 
  \sum_{\vec{r} = \vec{p}\uplus \vec{q}} \G(\vec{r};x) \; , 
\label{eq:halgebra} \end{eqnarray} 
where $\vec p, \vec q$ stand for the $p$ and $q$ components of the indices 
of the two 2dHPLs, while $\vec{p}\uplus \vec{q}$ represents all possible 
mergers of $\vec{p}$ and $\vec{q}$ into the vector $\vec{r}$ with $r$ 
components, in which the relative orders of the elements of $\vec{p}$ 
and $\vec{q}$ are preserved. 

Another class of identities is obtained by integrating (\ref{eq:defh0}) 
by parts. These integration-by-parts (IBP) identities read:
\begin{eqnarray}
\G(m_1,\ldots,m_q;x) &=&  \G(m_1;x)\G(m_2,\ldots,m_q;x)
                        -\G(m_2,m_1;x)\G(m_3,\ldots,m_q;x) \nonumber \\
&& + \ldots + (-1)^{q+1} \G(m_q,\ldots,m_1;x)\;.
\label{eq:ibp}
\end{eqnarray} 
These identities are not fully linearly 
independent from the product identities.

By using Eq.\ (\ref{eq:halgebra}) at weight $w=2$ for 
all possible independent values of the 
indices, one obtains 10 independent relations between the 16 2dHPLs  
of weight 2 and products of 2 2dHPLs  of weight 1; those relations 
can be used for expressing 10 of the 2dHPLs  of weight 2 in terms of 6 2dHPLs  
of weight 2 and products of 2 2dHPLs  of weight 1. 
The choice of the 6 2dHPLs  (referred to, in this context, as 
`irreducible') is by no means unique; by choosing as irreducible 2dHPLs  
of weight 2 the 6 functions $\G(0,1;y),~\G(0,1-z;y),~\G(0,-z;
y),~\G(1-z,1;y),~\G(-z,1;y),~\G(-z,1-z;y)$,
the reduction equations expressing the 10 `reducible' 2dHPLs  of weight 2 
in terms of the irreducible 2dHPLs  read 
\begin{eqnarray}%
 \G(-z,0;y) &=& 
      \G(0;y) \G(-z;y)
     - \G(0,-z;y)\;, \nonumber \\ 
\G(1-z,0;y) &=& 
      \G(0;y) \G(1-z;y)
     - \G(0,1-z;y)\;, \nonumber \\ 
\G(1,0;y) &=& 
      \G(0;y) \G(1;y)
     - \G(0,1;y)\;, \nonumber \\ 
\G(-z,-z;y) &=& 
      \frac{1}{2} \G(-z;y) \G(-z;y)\;, \nonumber \\
 \G(1-z,-z;y) &=& 
      \G(1-z;y) \G(-z;y)
     - \G(-z,1-z;y) \;,\nonumber \\ 
\G(1,-z;y) &=& 
      \G(1;y) \G(-z;y)
     - \G(-z,1;y)\;, \nonumber \\ 
\G(1-z,1-z;y) &=& 
      \frac{1}{2} \G(1-z;y) \G(1-z;y)\;, \nonumber \\ 
\G(1,1-z;y) &=& 
      \G(1;y) \G(1-z;y)
     - \G(1-z,1;y)\;, \nonumber \\ 
\G(1,1;y) &=& 
      \frac{1}{2} \G(1;y) \G(1;y)\;, \nonumber \\ 
\G(0,0;y) &=& 
      \frac{1}{2} \G(0;y) \G(0;y) \;.
\label{eq:red2} 
\end{eqnarray} 
Similarly, at weight 3, one has 64 2dHPLs  and 44 independent product 
and IBP
identities, expressing 44 reducible 2dHPLs  in terms of 20 irreducible ones; 
at weight 4 there are 256 2dHPLs, 196 independent identities, 
and correspondingly 196 reducible and 60 irreducible 2dHPLs. 
\begin{table}[t]
{\footnotesize
\begin{center}
\vspace{0mm}
\begin{tabular}{|c|c|c|}\hline
\rule[0mm]{0mm}{3mm}
Weight & Indices & 2dHPLs  \\[1mm] \hline 
\rule[0mm]{0mm}{3mm}
$w=1$ & &  $\G(0;y)$; $\G(1;y)$; $\G(1-z;y)$; $\G(-z;y)$\\[1mm] 
\hline 
\rule[0mm]{0mm}{3mm}
$w=2$ & $(0,1)$ & $\G(0,1;y)$ \\[1mm] 
\cline{2-3} 
\rule[0mm]{0mm}{3mm}
      & $(0,1-z)$ & $\G(0,1-z;y)$\\[1mm] 
\cline{2-3} 
\rule[0mm]{0mm}{3mm}
      & $(0,-z)$ & $\G(0,-z;y)$ \\[1mm] 
\cline{2-3} 
\rule[0mm]{0mm}{3mm}
      & $(0,1-z,1)$ & $\G(1-z,1;y)$\\[1mm] 
\cline{2-3} 
\rule[0mm]{0mm}{3mm}
      & $(0,-z,1)$ & $\G(-z,1;y)$\\[1mm] 
\cline{2-3} 
\rule[0mm]{0mm}{3mm}
      & $(0,-z,1-z)$ & $\G(-z,1-z;y)$\\[1mm] 
\hline 
\rule[0mm]{0mm}{3mm}
$w=3$ & $(0,1)$ & 
$\G(0,0,1;y)$; $\G(0,1,1;y)$\\[1mm] 
\cline{2-3} 
\rule[0mm]{0mm}{3mm}
      & $(0,1-z)$ & 
$\G(0,0,1-z;y)$; $\G(0,1-z,1-z;y)$\\[1mm] 
\cline{2-3} 
\rule[0mm]{0mm}{3mm}
      & $(0,-z)$ & 
$\G(0,0,-z;y)$; $\G(0,-z,-z;y)$\\[1mm] 
\cline{2-3} 
\rule[0mm]{0mm}{3mm}
      & $(0,1-z,1)$ & 
$\G(0,1-z,1;y)$; $\G(1-z,0,1;y)$;
$\G(1-z,1,1;y)$;
$\G(1-z,1-z,1;y)$\\[1mm] 
\cline{2-3} 
\rule[0mm]{0mm}{3mm}
      & $(0,-z,1)$ & 
$\G(0,-z,1;y)$; $\G(-z,0,1;y)$;
$\G(-z,1,1;y)$;
$\G(-z,-z,1;y)$\\[1mm] 
\cline{2-3} 
\rule[0mm]{0mm}{3mm}
      & $(0,-z,1-z)$ & $\G(0,-z,1-z;y)$;
                       $\G(-z,0,1-z;y)$;
$\G(-z,1-z,1-z;y)$;
$\G(-z,-z,1-z;y)$
\\[1mm] 
\cline{2-3}
\rule[0mm]{0mm}{3mm}
       & $(0,-z,1-z,1)$ &     
$\G(1-z,-z,1;y)$;
$\G(-z,1-z,1;y)$
\\[1mm] 
\hline 
\rule[0mm]{0mm}{3mm}
$w=4$ & $(0,1)$ & 
$\G(0,0,0,1;y)$; $\G(0,0,1,1;y)$; $\G(0,1,1,1;y)$\\[1mm] 
\cline{2-3} 
\rule[0mm]{0mm}{3mm}
      & $(0,1-z)$ & 
$\G(0,0,0,1-z;y)$; $\G(0,0,1-z,1-z;y)$; $\G(0,1-z,1-z,1-z;y)$\\[1mm] 
\cline{2-3} 
\rule[0mm]{0mm}{3mm}
      & $(0,-z)$ & 
$\G(0,0,0,-z;y)$; $\G(0,0,-z,-z;y)$; $\G(0,-z,-z,-z;y)$\\[1mm] 
\cline{2-3} 
\rule[0mm]{0mm}{3mm}
      & $(0,1-z,1)$ & 
$\G(0,0,1-z,1;y)$;
$\G(0,1,1-z,1;y)$;
$\G(0,1-z,0,1;y)$; \\[1mm] &&
$\G(0,1-z,1,1;y)$;
$\G(0,1-z,1-z,1;y)$;
$\G(1-z,0,0,1;y)$; \\[1mm] &&
$\G(1-z,0,1,1;y)$;
$\G(1-z,0,1-z,1;y)$;
$\G(1-z,1,1,1;y)$; \\[1mm] &&
$\G(1-z,1-z,0,1;y)$;
$\G(1-z,1-z,1,1;y)$;
$\G(1-z,1-z,1-z,1;y)$\\[1mm]
\cline{2-3} 
\rule[0mm]{0mm}{3mm}
      & $(0,-z,1)$ & 
$\G(0,0,-z,1;y)$;
$\G(0,1,-z,1;y)$;
$\G(0,-z,0,1;y)$; \\[1mm] &&
$\G(0,-z,1,1;y)$;
$\G(0,-z,-z,1;y)$;
$\G(-z,0,0,1;y)$; \\[1mm] &&
$\G(-z,0,1,1;y)$;
$\G(-z,0,-z,1;y)$;
$\G(-z,1,1,1;y)$; \\[1mm] &&
$\G(-z,-z,0,1;y)$;
$\G(-z,-z,1,1;y)$;
$\G(-z,-z,-z,1;y)$\\[1mm]
\cline{2-3} 
\rule[0mm]{0mm}{3mm}
      & $(0,-z,1-z)$ & 
$\G(0,0,-z,1-z;y)$;
$\G(0,1-z,-z,1-z;y)$;
$\G(0,-z,0,1-z;y)$; \\[1mm] &&
$\G(0,-z,1-z,1-z;y)$;
$\G(0,-z,-z,1-z;y)$;
$\G(-z,0,0,1-z;y)$; \\[1mm] &&
$\G(-z,0,1-z,1-z;y)$;
$\G(-z,0,-z,1-z;y)$;
$\G(-z,1-z,1-z,1-z;y)$; \\[1mm] &&
$\G(-z,-z,0,1-z;y)$;
$\G(-z,-z,1-z,1-z;y)$;
$\G(-z,-z,-z,1-z;y)$\\[1mm]
\cline{2-3} 
\rule[0mm]{0mm}{3mm}
      & $(0,-z,1-z,1)$ & 
$\G(0,1-z,-z,1;y)$;
$\G(0,-z,1-z,1;y)$;
$\G(1-z,0,-z,1;y)$;\\[1mm] &&
$\G(1-z,1-z,-z,1;y)$;
$\G(1-z,-z,0,1;y)$;
$\G(1-z,-z,1,1;y)$;\\[1mm] &&
$\G(1-z,-z,1-z,1;y)$;
$\G(1-z,-z,-z,1;y)$;
$\G(-z,0,1-z,1;y)$;\\[1mm] &&
$\G(-z,0,-z,1;y)$;
$\G(-z,1,1-z,1;y)$;
$\G(-z,1-z,0,1;y)$;\\[1mm] &&
$\G(-z,1-z,1,1;y)$;
$\G(-z,1-z,1-z,1;y)$;
$\G(-z,1-z,-z,1;y)$;\\[1mm] &&
$\G(-z,-z,1-z,1;y)$ \\[1mm] \hline
\end{tabular}
\caption{List of irreducible 2dHPLs  chosen 
in the  numerical implementation.}
\label{tab:irred}
\end{center}}
\end{table}

The 2dHPLs chosen as the irreducible set in the program described here 
are listed in Table~\ref{tab:irred}; for convenience of later use 
(see in particular Section~\ref{sec:numer}) we grouped them 
according to the different combinations of the occurring indices. 
Note in particular that, with our choice, the index $(0)$ is never present 
as rightmost (or trailing) index of the 2dHPLs of the irreducible 
set (except for the trivial case, at $w=1$, of $\H(0;y)=\ln{y}$). 

\section{Special values of the argument}
\label{sec:values} 
\setcounter{equation}{0} 
At $ y=1-z$, $\ y=-z$ and $\ y=1$, the 2dHPLs can be expressed in terms 
of HPLs of argument $z$.  
\par 
To obtain such expressions, one can write the obvious equation 
\begin{equation}
\G(\vec{m}(z);y(z)) = \G(\vec{m}(z=0);y(z=0)) + \int_0^z \d z^{\prime}
        \frac{\d}{\d z^{\prime}} \G(\vec{m}(z^{\prime});y(z^{\prime}))\; ,
\end{equation} 
where $y(z)$ stands for any of the above particular values $ y=1-z $, 
$y=-z$ or $y=1$ (the argument applies for $y$ taking the constant value $1$ 
as well), $\vec{m}(z)$ is any set of indices and it is understood that 
the boundary $z=0$ is to be replaced by $z=1$ if $\G(\vec{m}(z=0);y(z=0))$ 
is divergent. The derivative $\d/\d{z^{\prime}}$ is then 
carried out on the representation of $\G(\vec{m}(z^{\prime});y(z^{\prime}))$ 
as a repeated integral (\ref{eq:zlimone}) or (\ref{eq:zlimtwo}). 
Quadratic factors of the form $[\g(m_i;t_j)]^2$, which can appear when 
carrying out the $z^{\prime}$-derivative explicitly, are reduced to 
single powers by integration by parts and partial fractioning (to be 
iterated when needed). The resulting repeated integrals can then be
identified as a combination of multiple integral representations of 
HPLs of argument $z$. 

As an example, we evaluate 
$ \G(1,1-z;y) $ in $y=1-z$: 
\begin{eqnarray}
\G(1,1-z;1-z) & = & \G(1,0;0) + \int_1^z \d z^{\prime} 
                    \frac{\d }{\d z^{\prime}} \G(1,1-z^{\prime};1-z^{\prime})
                 \nonumber \\
&=& \int_1^z \d z^{\prime} \frac{\d }{\d z^{\prime}} \left[ 
                   \int_0^{1-z^{\prime}} \frac{\d t_1}{t_1-1}
                   \int_0^{t_1} \frac{\d t_2}{t_2+z^{\prime}-1} \right] 
                 \nonumber \\
&=& \int_1^z \d z^{\prime} \left[ \frac{1}{z^{\prime}} 
    \int_0^{1-z^{\prime}} \frac{\d t_2}{t_2+z^{\prime}-1} 
  - \int_0^{1-z^{\prime}} \frac{\d t_1}{t_1-1} \int_0^{t_1} 
                          \frac{\d t_2}{(t_2+z^{\prime}-1)^2}\right]
                 \nonumber \\
&=& \int_1^z \d z^{\prime} \left(- \frac{1}{1-z^{\prime}}
                               -\frac{1}{z^{\prime}}\right) 
                     \int_0^{1-z^{\prime}} \frac{\d t_1}{1-t_1} 
                 \nonumber \\
&=& \int_1^z \d z^{\prime} \left(- \frac{1}{1-z^{\prime}}
                            -\frac{1}{z^{\prime}}\right)  \H(1;1-z^{\prime})
                 \nonumber \\
&=& \int_1^z \d z^{\prime} \left(+ \frac{1}{1-z^{\prime}}
                        +\frac{1}{z^{\prime}}\right)  \H(0;z^{\prime})
                 \nonumber \\
&=& \H(1,0;z) - \H(1,0;1) + \H(0,0;z) \; . 
\end{eqnarray} 
Note the appearance, in the above calculation, of $\H(1;1-z^{\prime}) $, 
which is then replaced by $ -H(0;z^{\prime}) $ (according to the 
very definition, see for instance (\ref{eq:levelone})). In more 
complicated cases the replacement is carried out in a recursive manner, 
\ie\ by using in the transformation of 2dHPLs of weight $w$ 
the results obtained previously for 2dHPLs of weight less than $w$. 
\par 
The above sketched algorithm has been programmed in \FORM~\cite{form} 
to derive the transformation formulae in $y=1-z$, $\ y=-z$ and $\ y=1$ for
all the irreducible 2dHPLs up to weight 4. 

\section{Identities for related arguments} 
\label{sec:ids} 
\setcounter{equation}{0} 
In the context of the numerical evaluation of the 2dHPLs, it is sometimes 
convenient to map $\G(\vec{m}(z);y)$ into $\G(\vec{m}(z);y^\prime)$ with 
$y^\prime = 1-z-y$. Indeed, this mapping allows us
to rewrite a 2dHPL whose 
argument $y$ lies in the range $(1-z)/2 < y \leq (1-z)$ in terms of 
2dHPLs with argument $y^\prime$ in the range 
$0\leq y^\prime \leq (1-z)/2$. As a consequence, power series expansions 
for the 2dHPLs are needed only in the region $0\leq y \leq (1-z)/2$,
thus avoiding potentially non-analytic points at $y=(1-z)$.
\par 
Much as in the previous section, the mapping can be obtained by writing 
the obvious equation 
\begin{equation}
\G(\vec{m}(z);1-y-z) = \G(\vec{m}(z);1-z) + \int_0^y \d y^{\prime}
                     \frac{\d}{\d y^{\prime}} \G(\vec{m}(z);1-y^\prime-z)\; ,
\end{equation} 
where the boundary $y=0$ can be replaced by $y=1-z \ $ if 
$\G(\vec{m}(z);1-z)$ is divergent (as  is the case when $1-z$ 
appears as leftmost index in $\vec{m}(z)$). 
The $y^\prime$-derivative is again carried out on the 
multiple integral representation of $\G(\vec{m}(z);1-y-z)$ 
(\ref{eq:zlimone}),(\ref{eq:zlimtwo}). A repeated application of 
integration by parts and partial fractioning then generates an 
expression that can be identified as a bilinear combination of 
2dHPLs $\G(\vec{m}(z);y)$ and ordinary HPL $\H(\vec{m};z)$. 
As an example, consider 
\begin{eqnarray} 
\G(1,1-z;1-y-z) & = 
& \G(1,1-z;1-z) + \int_0^y \d y^{\prime} 
\frac{\d}{\d y^{\prime}} \G(1,1-z;1-y^\prime-z)   \nonumber \\ 
&=& \G(1,1-z;1-z) + \int_0^y \d y^{\prime} 
\frac{\d}{\d y^{\prime}} \left[ \int_0^{1-y^\prime-z}\frac{\d t_1}{t_1-1} 
\int_0^{t_1}\frac{\d t_2}{t_2+z-1} \right]        \nonumber \\ 
&=& \G(1,1-z;1-z) + \int_0^y \d y^{\prime} \frac{-1}{-y^\prime-z} 
 \int_0^{1-y^\prime-z}\frac{\d t_2}{t_2+z-1}      \nonumber \\ 
&=& \G(1,1-z;1-z) + \int_0^y \frac{\d y^{\prime}}{y^\prime+z} 
           \left[ \G(0;y^\prime) + \H(1;z)\right] \nonumber \\ 
&=& \H(1,0;z) - \H(0,1;1) + \H(0,0;z) + \G(-z,0;y) + \G(-z;y)\H(1;z) \; . 
\end{eqnarray} 
Note the use, in the last step, of the results of the previous section. 
Interchange formulae $ y \to 1-y-z $ have been derived for 2dHPLs up to 
weight 4 by programming the above sketched algorithm in \FORM~\cite{form}. 
As the
transformation algorithm presented above, the interchange algorithm also 
works recursively, by using the
interchange formulae obtained previously for lower
weights. 

\section{The analyticity properties} 
\label{sec:anal} 
\setcounter{equation}{0} 

Let us recall that we are interested only in the region 
$0\le y \le 1$, $\ \ 0\le z \le 1$, $\ \ 0\le y+z \le 1$. 
There are two possible right cuts in $y$, 
\begin{itemize} 
\item $y=1$, coming from $\g(1;y) = 1/(y-1)$ (the same as for the HPLs), 
\item $y=1-z$, coming from $\g(1-z;y) = 1/(y+z-1)$, 
\end{itemize} 
and a single left cut 
\begin{itemize} 
\item $y=-z$, coming from $\g(-z;y) = 1/(y+z)$  (remember $z\ge 0$) 
\end{itemize} 
\par 
We use a set of irreducible 2dHPLs in which the indices appear in the order 
$(0,(-z),(1-z),1)$, so that a rightmost $(-z)$ index can have at its left 
only $(0)$'s or $(-z)$'s, a rightmost $(1-z)$ can have $(0)$'s, $(-z)$'s or 
$(1-z)$'s, and finally a rightmost $(1)$ can have at its left 
$(0)$'s, $(-z)$'s, 
$(1-z)$'s or $(1)$'s. As a consequence, a rightmost $(0)$ and the related 
logarithmic cut at $y=0$ are never present in the 2dHPLs of the 
irreducible set chosen here (except in the trivial case, at $w=1$, of 
$\H(0;y) = \ln\, y $). 
 
Following \cite{1dhpl}, each 2dHPL can be written as 
\begin{equation} 
 \G(\vec m(z);y) = \Gp(\vec m(z);y) + \Gm(\vec m(z);y) \; , 
\label{eq:rlsplit} 
\end{equation} 
where $\Gp(\vec m(z);y)$ contains only one or both right cuts, and 
$\Gm(\vec m(z);y)$ contains only the left cut. 
Note that we keep together the two right cuts, when both are present. 

The separation of the cuts is carried out iteratively. At weight $w=1$ 
the cut-structure is 
\begin{eqnarray} 
\G(1;y) &=& \Gp(1;y) \; ,     \nonumber \\ 
\G(1-z;y) &=& \Gp(1-z;y) \; , \nonumber \\ 
\G(-z;y) &=& \Gm(-z;y) \;, 
\label{eq:primary} 
\end{eqnarray} 
as $\Gm(1;y) = \Gm(1-z;y) = \Gp(-z;y) = 0$. \par 
At higher weights, the following separation formulae apply: 
\begin{equation} 
\G(a(z),\vec{m}(z);y) = \Gp(a(z),\vec{m}(z);y) + \Gm(a(z),\vec{m}(z);y) \; , 
\label{eq:sepcut}
\end{equation} 
with 
\begin{eqnarray} 
\G_{\pm}(0,\vec{m}(z);y) &=& \int_0^y \frac{\d y^\prime}{y^\prime} 
\G_{\pm}(\vec{m}(z);y^\prime) \; ,\nonumber \\
\Gp(1,\vec{m}(z);y) &=& \int_0^y \frac{\d y^\prime}{y^\prime-1} 
\left(\Gp(\vec{m}(z);y^\prime) + \Gm(\vec{m}(z);1) \right) \; ,\nonumber \\
\Gm(1,\vec{m}(z);y) &=& \int_0^y \frac{\d y^\prime}{y^\prime-1} 
\left(\Gm(\vec{m}(z);y^\prime)-\Gm(\vec{m}(z);1)\right) \; ,\nonumber \\
\Gp(1-z,\vec{m}(z);y) &=& \int_0^y \frac{\d y^\prime}{y^\prime-1+z} 
\left(\Gp(\vec{m}(z);y^\prime) + \Gm(\vec{m}(z);1-z) \right) \; ,\nonumber \\
\Gm(1-z,\vec{m}(z);y) &=& \int_0^y \frac{\d y^\prime}{y^\prime-1+z} 
\left(\Gm(\vec{m}(z);y^\prime)-\Gm(\vec{m}(z);1-z)\right) \; ,\nonumber \\
\Gp(-z,\vec{m}(z);y) &=& \int_0^y \frac{\d y^\prime}{y^\prime+z} 
\left(\Gp(\vec{m}(z);y^\prime) - \Gp(\vec{m}(z);-z) \right) \; ,\nonumber \\
\Gm(-z,\vec{m}(z);y) &=& \int_0^y \frac{\d y^\prime}{y^\prime+z} 
\left(\Gm(\vec{m}(z);y^\prime)+\Gp(\vec{m}(z);-z)\right) \; .
\label{eq:sepcuts}
\end{eqnarray}
The cut corresponding to the rightmost index, see~(\ref{eq:primary}), is 
called the primary cut, the other cut (when present) the secondary cut. 
The separation of cuts does require the computation of the 2dHPLs 
at the points $ y=1-z $, $y=-z$ or $y=1$, which was explained in 
Section~\ref{sec:values}.

It is important to note that, for a given 2dHPL $\G(\vec{m}(z);y)$,
the secondary cut consists of 
a product of $\H(\vec{a};z)$ times only 2dHPLs of weight lower than 
$\G(\vec{m}(z);y)$. If the 2dHPLs are evaluated for increasing weight, the 
secondary cut contribution does not need to be evaluated anew. 
Only the principal cut contains 2dHPLs of the same 
weight as $\G(\vec{m}(z);y)$, and has to be evaluated. 
\par 
As an example of the separation of the cuts according to the rules of 
(\ref{eq:sepcuts}), let us consider 
\begin{equation} 
  \G(0,-z,1,1;y) = \Gp(0,-z,1,1;y) + \Gm(0,-z,1,1;y)\; , \nonumber \\ 
\end{equation} 
with 
\begin{eqnarray} 
  \Gp(0,-z,1,1;y) &=& \G(0,-z,1,1,y) - \H(-1,-1;z)\G(0,-z;y)\;, \nonumber \\ 
  \Gm(0,-z,1,1;y) &=& \H(-1,-1,z)\G(0,-z,y) \; . 
\end{eqnarray} 

\section{The numerical evaluation} 
\label{sec:numer} 
\setcounter{equation}{0} 
As in~\cite{1dhpl}, we evaluate the irreducible 2dHPLs first, and then 
the reducible ones by using the formulae expressing them in terms of 
the irreducible ones, see Section~\ref{sec:alg}. \par
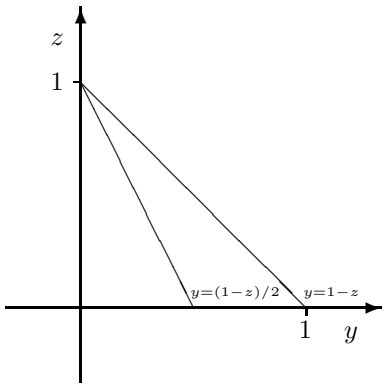
\begin{figure}[b] 
\begin{center} 
\begin{picture}(6,6) 
\thicklines 
\put(0,1){\vector(1,0){5}} 
\put(1,0){\vector(0,1){5}} 
\thinlines 
\put(1,4){\line(1,-1){3}} 
\put(1,4){\line(1,-2){1.5}} 
\put(1,4){\line(-1,0){0.1}} 
\put(4,1){\line(0,-1){0.1}} 
\put(0.6,4.5){$z$} 
\put(4.5,0.6){$y$} 
\put(0.6,3.9){$1$} 
\put(3.9,0.6){$1$} 
\put(3.85,1.1){\makebox(0,0)[l]{$\,^{\,^{y=1-z}}$}} 
\put(2.35,1.1){\makebox(0,0)[l]{$\,^{\,^{y=(1-z)/2}}$}} 
\end{picture} 
\end{center} 
\caption{Kinematic regions for the evaluation of the 2dHPLs} 
\label{fig:kin} 
\end{figure}

 
For the purpose of the numerical evaluation of the 2dHPLs, we 
restrict ourselves to the following values of the arguments,
depicted by the outer triangle in Fig.~\ref{fig:kin}:
\begin{equation} 
       0\leq y \leq 1\;, \qquad 0\leq z \leq 1\;, 
                         \quad \mbox{with} \quad y+z\leq 1\;, 
\end{equation} 
which is the only kinematical region relevant to the calculation of 
physical amplitudes in quantum field theory (note that there are other 
kinematically allowed regions in the $(y,z)$-plane, which can however 
be related to the above triangle by analytic continuation, see 
Appendix of~\cite{doublebox}). 

When the vector of the indices of the 2dHPL contains only a single 
index different from $(0)$, which can however be repeated several 
times, one can rescale the argument, obtaining in that way an HPL of 
suitable argument, which can be evaluated with the routine 
{\tt hplog}~\cite{1dhpl}. 

If the argument $y$ is in the inner triangle on the right of 
Fig.~\ref{fig:kin}, $\ (1-z)/2 < y \le (1-z) $, we use 
the results of Section~\ref{sec:ids} to express the 2dHPLs in terms 
of 2dHPLs of argument $y'=(1-y-z)$, with $0\le y' \le (1-z)/2 $. 

Finally, when the argument $y$ is in the triangle $0\le y \le (1-z)/2 $ , 
on the left of Fig.~\ref{fig:kin}, we evaluate 
the primary cuts of the 2dHPLs of the irreducible set by their series 
expansion in powers of $y$ around $y=0$, their secondary cuts by their 
expression in terms of 2dHPLs of lower weight, see 
(\ref{eq:sepcut}) and (\ref{eq:sepcuts}), and then sum the two cuts. 

The coefficients of the expansions in $y$ of the primary cuts are in turn 
functions of $z$. 
For $0\le z < 1/2$ it is convenient to expand the coefficients, which often 
show nominal divergences in $1/z$, generating serious numerical instabilities, 
in powers of $z$, obtaining in that way a stable and quickly convergent 
expression. For some of the 2dHPLs,  these $z$-dependent coefficients 
can contain HPLs of $z$, which are non-analytic at $z=1$, thus resulting in 
a failure of the power series expansion. In these cases, 
for $1/2 \le z \le 1$, the exact expressions are used. 

All the various expansions (in $y$, but occasionally also in $z$) can be 
accelerated by the Bernoulli-like changes of variables~\cite{btrans} 
already used in 
\cite{1dhpl}, and the resulting series can be further economized by standard 
use of Chebyshev polynomials. The expansion in $y$ is always performed first, 
yielding $z$-dependent coefficients. The {\tt FORTRAN} subroutine implementing 
the numerical evaluation of the $z$-dependent coefficients checks whether 
it is called repeatedly 
for the same value of $z$. In this case, the $z$-dependent expansion 
coefficients are not re-evaluated, which yields a considerable 
acceleration of running time (about a factor 3.5). 
\par
In the following subsections, we describe in detail the expansions used for 
the 2dHPLs, depending on the combination of indices appearing in the
2dHPL under consideration. All these expansions have been generated using 
\FORM~\cite{form}, whose output was then converted to {\tt FORTRAN} input of 
the required precision with a dedicated {\tt C} program, 
rewriting in particular the exact coefficients generated by {\tt FORM} 
as double precision floating-point numbers. 

\subsection{The indices $(0,1)$} 
If only the indices $(0)$ and $(1)$ appear in $\vec{m}$, the 2dHPLs 
$\ \G(\vec{m};y)$ are (up to an overall sign, which is $+1$ for an even 
number of $(1)$ in $\vec{m}$ and $-1$ for an odd number of $(1)$ in 
$\vec{m}$) equal to the HPLs $\ \H(\vec{m};y)$, to be evaluated by means 
of {\tt hplog}~\cite{1dhpl}. 

\subsection{The indices $(0,1-z)$} 
If only $(0)$ and $(1-z)$ appear in $\vec{m}(z)$, one can perform the 
mere change of variable from $t_i$ to $t^\prime_i=t_i/(1-z)$ in the 
multiple integral representation (\ref{eq:zlimone}),(\ref{eq:zlimtwo}). 
The individual integrands transform as follows:
\begin{equation}
\d t_i\, \g(0;t_i) \to \d t^\prime_i \,\f(0;t^\prime_i) \; , \qquad 
\d t_i\,  \g(1-z;t_i) \to - \d t^\prime_i \, \f(1;t^\prime_i) \; , 
\end{equation} 
and the 2dHPLs are re-expressed as HPLs of argument $y/(1-z)$, 
which can then be evaluated by means of {\tt hplog}~\cite{1dhpl}. 

As an example, consider 
\begin{eqnarray} 
\G(0,1-z,1-z;y) &=& \int_0^y \d t_1 \, \g(0;t_1) 
\int_0^{t_1} \d t_2 \, \g(1-z;t_2) 
\int_0^{t_2} \d t_3 \, \g(1-z;t_3) \nonumber \\ 
&=& (-1)^2 \int_0^{\frac{y}{1-z}} \d t^\prime_1 \, \f(0;t^\prime_1) 
\int_0^{t^\prime_1} \d t^\prime_2 \, \f(1;t^\prime_2) 
\int_0^{t^\prime_2} \d t^\prime_3 \, \f(1;t^\prime_3) \nonumber \\ 
&=& \H\left( 0,1,1;\frac{y}{1-z}\right)\; . 
\end{eqnarray} 

\subsection{The indices $(0,-z)$} 
If only $(0)$ and $(-z)$ appear in $\vec{m}(z)$, the 2dHPLs can again 
be re-expressed 
as HPLs, in this case  of argument $-y/z$ by a mere change of variable 
 from $t_i$ to $t^\prime_i=-t_i/z$ 
in the multiple integral representation, which transforms the integrands 
\begin{equation} 
\d t_i\, \g(0;t_i) \to \d t^\prime_i \,\f(0;t^\prime_i) \; , \qquad 
\d t_i\, \g(-z;t_i) \to - \d t^\prime_i \, \f(1;t^\prime_i) \; . 
\end{equation} 
As a result, one obtains HPLs containing only the indices (0,1) with argument 
$\xi=-y/z$, which can then be evaluated by means of {\tt hplog}~\cite{1dhpl}. 
It is apparent that this transformation can be applied 
unambiguously only if the trailing (or right-most) index is different from 
(0), which is however always the case for the 2dHPLs in the irreducible set. 

As an example, we evaluate 
\begin{eqnarray} 
\G(0,0,-z;y) &=& \int_0^y \d t_1 \, \g(0;t_1) 
\int_0^{t_1} \d t_2 \, \g(0;t_2) 
\int_0^{t_2} \d t_3 \, \g(-z;t_3) \nonumber \\ 
&=& (-1) \int_0^{-\frac{y}{z}} \d t^\prime_1 \, \f(0;t^\prime_1) 
\int_0^{t^\prime_1} \d t^\prime_2 \, \f(0;t^\prime_2) 
\int_0^{t^\prime_2} \d t^\prime_3 \, \f(1;t^\prime_3) \nonumber \\ 
&=& -\H\left( 0,0,1;-\frac{y}{z}\right)\; . 
\end{eqnarray} 

\subsection{The indices $(0,1-z,1)$} 

If only the indices  $(0,1-z,1)$ or $(1-z,1)$
are present in $\vec m(z)$, then 
$\G(\vec{m}(z);y)$ contains only the two right cuts in the complex plane,
but no left cut. The separation of cuts is therefore not carried out. 
These 2dHPLs are expanded in $s=-\log(1-y/(1-z))$, which corresponds 
to the Bernoulli-like variable appropriate to the cut 
at $y=(1-z)$, always closer to the origin $y=0$ than the cut at $y=1$. 
As $ 0 \leq y \leq (1-z)/2 $, so that $ s \leq \ln{2} \approx 0.6931\ldots$, 
we rescale $s$ by a factor $4/3$; in so doing, one still has 
$(4s/3)^n \leq 1 $ on the whole interval $0 \leq y \leq (1-z)/2\ $. 
The coefficients of the terms in $(4s/3)^n$, which are 
smaller by a factor of $(3/4)^n$ after rescaling, are finite-order 
polynomials in $z$, well-behaved in the whole interval $0\leq z\leq 1\ $. The 
convergence of the $z$-expansion is further improved by rewriting the 
resulting polynomials in $z$ in terms of Chebyshev polynomials 
in $(2z-1)$, which finally allows coefficients 
smaller than the desired accuracy of the numerical evaluation
to be identified and dropped. 

For example, the expansion of $\G(0,0,1-z,1;y)$ reads
\begin{eqnarray}
\lefteqn{\G(0,0,1-z,1;y)= }\nonumber \\    
&&       + \left(\frac{4s}{3}\right)^2 \left(
          + \frac{9}{256}
          - \frac{9}{256}T_1(2z-1)
         \right)
       + \left(\frac{4s}{3}\right)^3 \left(
          - \frac{1}{64}
          + \frac{15}{1024}T_1(2z-1)
          + \frac{1}{1024}T_2(2z-1)
         \right)
\nonumber\\ &&
       + \left(\frac{4s}{3}\right)^4 \left(
          + \frac{441}{131072}
          - \frac{1485}{524288}T_1(2z-1)
          - \frac{63}{131072}T_2(2z-1)
          - \frac{27}{524288}T_3(2z-1)
         \right)
\nonumber\\ &&
       + \left(\frac{4s}{3}\right)^5 \bigg(
          - \frac{25749}{65536000}
          + \frac{6831}{26214400}T_1(2z-1)
          + \frac{3303}{32768000}T_2(2z-1)
          + \frac{729}{26214400}T_3(2z-1)
\nonumber \\ &&
          + \frac{243}{65536000}T_4(2z-1)
         \bigg)
       + \left(\frac{4s}{3}\right)^6 \bigg(
          + \frac{7083}{419430400}
          + \frac{27}{83886080}T_1(2z-1)
          - \frac{459}{52428800}T_2(2z-1)
\nonumber \\ &&
          - \frac{63}{10485760}T_3(2z-1)
          - \frac{891}{419430400}T_4(2z-1)
          - \frac{27}{83886080}T_5(2z-1)
         \bigg)
       + {\cal O} \left(s^7\right)
\end{eqnarray}
The presence of two non-separable right-hand cuts in  functions 
bearing this combination of indices implies 
a somewhat slow convergence of the power series expansion. To attain 
the aimed double precision accuracy of 
$3\times 10^{-15}$, one must keep 18 terms in the $s$-expansion of 
$\G(0,0,1-z,1;y)$ and expand the coefficients up to the 15th Chebyshev 
polynomial. Other functions with the same combination of indices 
require up to 21 terms in the expansion, and up to the 19th Chebyshev
polynomial in $(2z-1)$.

\subsection{The indices $(0,-z,1)$}
If the indices  $(0,-z,1)$ or only $(-z,1)$ 
are present in $\vec m(z)$, then
$\G(\vec{m}(z);y)$ contains two cuts, one left cut in $y=-z$ and 
one right cut in $y=1$. 
With this combination of indices, the irreducible 2dHPLs are 
chosen to always contain $(1)$ as rightmost index. The primary cut 
of these functions is therefore always the right cut in $y=1$, 
with $y=-z$ appearing only as secondary cut. 

After separating the left and right cuts according to (\ref{eq:rlsplit}),
only $\G_+(\vec{a};y)$ needs to be evaluated, since the secondary 
$\G_-(\vec{a};y)$ is always expressed in terms of known functions of lower 
weight; $\G_+(\vec{a};y)$ is expanded in $y$ at $y=0$ and $y$ is then replaced 
by the Bernoulli-type variable $r=-\log(1-y)$. The expansion coefficient of 
$r^n$ is a function of $z$, which contains an overall factor $z^{-n}$ 
multiplying a numerator containing a finite order polynomial in $z$ 
plus further $n$-th order polynomials in $z$ times HPLs of argument $z$ 
and indices of the subset $(0,-1)$ only. In $z\to 0$, the coefficient 
and all its derivatives in $z$ are finite. The coefficients are expanded in 
$v=H(-1;z)$, which is the Bernoulli-type variable appropriate here. 
The convergence of the resulting series in $v$ is further improved by 
re-expressing them in terms of Chebyshev polynomials, with $v$ rescaled 
appropriately as ($8v/3-1$). 

For example, the expansion of $\G(0,1,-z,1;y)$ in powers of $r$ reads 
\begin{eqnarray} 
\lefteqn{\G(0,1,-z,1;y)  }\nonumber \\ 
&= & \G_+(0,1,-z,1;y) + \G_-(0,1,-z,1;y)\nonumber \\ 
&=& + \left(\frac{r}{z}\right) \biggl( - 2z\H(-1,-1;z) \biggr) 
       + \left(\frac{r}{z}\right)^2 \left(
          + \frac{z}{4}\H(-1;z)
          +  \frac{z^2}{2}\H(-1,-1;z)
         \right)\nonumber \\
& & 
       + \left(\frac{r}{z}\right)^3 \left(
          +  \frac{z^2}{18}
          -  \frac{z}{18}\H(-1;z)
          -  \frac{5z^2}{36}\H(-1;z)
          -  \frac{z^3}{18}\H(-1,-1;z)
         \right)\nonumber \\
& & 
       + \left(\frac{r}{z}\right)^4 \left(
          -  \frac{z^2}{48}
          -  \frac{z^3}{24}
          +  \frac{z}{48}\H(-1;z)
          +  \frac{5z^2}{96}\H(-1;z)
          +  \frac{z^3}{24}\H(-1;z)
         \right)\nonumber \\
& & 
       + \left(\frac{r}{z}\right)^5 \bigg(
          +  \frac{z^2}{100}
          +  \frac{7z^3}{300}
          +  \frac{29z^4}{1800}
          -  \frac{z}{100}\H(-1;z)
          -  \frac{17z^2}{600}\H(-1;z)
          -  \frac{97z^3}{3600}\H(-1;z)
\nonumber \\
& & 
          -  \frac{31z^4}{3600}\H(-1;z)
          +  \frac{z^5}{1800}\H(-1,-1;z)
         \bigg) + {\cal O}(r^6)\nonumber \\
& &  
       + \H(-1;z)\G(0,1,-z;y)
          - 2\H(-1,-1;z)\G(0,1;y)\;.
\end{eqnarray}        
The subsequent expansion of the coefficients of the $r$-expansion 
in Chebyshev polynomials in the rescaled 
$v$ does yield rational coefficients, involving large integer numbers in 
numerator and denominator (recall that the Chebyshev expansion is 
performed on a power series, not on a finite order polynomial). 
For the sake of brevity, we shall only display the function truncated 
to order $r^2$ and $v^2$, which is sufficient to 
highlight the structure of the expansion: 
\begin{eqnarray} 
\lefteqn{\G(0,1,-z,1;y)  }\nonumber \\ 
&= & 
       + \left(\frac{4r}{3}\right) \left(
          - \frac{81}{512}
          - \frac{27}{128}T_1\left( \frac{8v}{3}-1\right)
          - \frac{27}{512}T_2\left( \frac{8v}{3}-1\right) 
          \right)\nonumber \\
&&
       + \left(\frac{4r}{3}\right)^2 \left(
          + \frac{4797}{32768}
          + \frac{135}{8192}T_1\left( \frac{8v}{3}-1\right)
          + \frac{351}{32768}T_2\left( \frac{8v}{3}-1\right) + {\cal O}(v^3)
          \right) + {\cal O}(r^3) \nonumber \\
&&
       + \H(-1;z)\G(0,1,-z;y)
          - 2\H(-1,-1;z)\G(0,1;y)\; .
\end{eqnarray}
To obtain the desired accuracy of 
$3\times 10^{-15}$, one must keep 17 terms in the $r$-expansion of 
$\G(0,1,-z,1;y)$ and expand the coefficients up to the 10th Chebyshev 
polynomial in the rescaled $v$. Other functions with the same combination 
of indices require up to 18 terms in $r$ and up to the 11th Chebyshev 
polynomial in rescaled 
$v$. Rewriting the expansion in $r$ in terms of Chebyshev polynomials 
yields no improvement as far as the number of arithmetic operations 
required in the evaluation of the functions is concerned: the number of 
operations to be performed does in fact get larger in most cases. 

\subsection{The indices $(0,-z,1-z)$}
If $\vec m(z)$  contains the indices  $(0,-z,1-z)$ or only $(-z,1-z)$, then
$\G(\vec{m}(z);y)$ contains two cuts, one left cut in $y=-z$ and 
one right cut in $y=1-z$. 
The irreducible 2dHPLs with this combination of indices are 
chosen to contain always a $(1-z)$ as rightmost index. Therefore, 
these functions have always the right cut in $y=1-z$ as primary cut, 
with $y=-z$ only appearing as secondary cut. 

After separating the left and right cuts according to (\ref{eq:rlsplit}),
only $\G_+(\vec{a};y)$ needs to be evaluated, since the secondary 
$\G_-(\vec{a};y)$ is always expressed in terms of known functions of 
lower weight; $\G_+(\vec{a};y)$ is expanded in $y$ at $y=0$, and $y$ is 
then replaced by the Bernoulli-type variable $s=-\log(1-y/(1-z))$. The 
coefficients of the $s$-expansion take a form very similar to the expansion 
coefficients of the 2dHPLs with indices $(0,1-z,1)$ described 
in the previous subsection. The coefficient of $s^n$ is 
indeed equal to an overall factor $z^{-n}$, 
multiplying a numerator containing a finite order polynomial 
plus further $n$-th order polynomials multiplying HPLs of argument $z$, 
which contain indices of the subset $(0,1)$ only. In $z\to 0$, the 
coefficient and all its derivatives in $z$ are finite. 

In contrast to the expansion 
coefficients described in the previous subsection, which are analytic 
in the whole interval $0\leq z \leq 1$, the coefficients in the 
$s$-expansion carried out here are finite, but non-analytic in the point 
$z=1$. It is therefore not possible to obtain for these coefficients 
a series representation valid over the whole interval $0\leq z \leq 1$. 
Instead, we cut this interval at $z=1/2$. For $0\leq z \leq 1/2$, 
we express $z$ in terms of
 $u=\H(1;z)$, which is the appropriate Bernoulli variable 
here, and expand the coefficients in $u$. Again, the convergence of the 
resulting series of powers of $u$ is further improved by re-expressing them 
in terms of Chebyshev polynomials, with $u$ rescaled appropriately. 
In the second interval $1/2 < z \leq 1$, 
we evaluate the coefficients using the exact expressions. To avoid 
large-scale numerical cancellations, we then compute the coefficients as 
a function of $1-z$ instead of $z$. 

As an example, we quote the $s$-expansion of $\G(-z,0,1-z,1-z;y)$:
\begin{eqnarray} 
\lefteqn{\G(-z,0,1-z,1-z;y)} \nonumber \\ 
&=& \G_+(-z,0,1-z,1-z;y) +\G_-(-z,0,1-z,1-z;y) \nonumber \\ 
&=& 
       + \left(\frac{s}{z}\right) \left( 
          + z \H(0,1,1;z) 
          - \H(0,1,1;z) 
          + z \H(1,1,1;z) 
          -   \H(1,1,1;z) 
          \right)\nonumber \\ 
&& 
       + \left(\frac{s}{z}\right)^2 \left(
          + \frac{1}{2}\H(0,1,1;z)
          - \frac{z}{2}\H(0,1,1;z)
          + \frac{1}{2}\H(1,1,1;z)
          - \frac{z}{2}\H(1,1,1;z)
          \right)\nonumber \\
&& 
       + \left(\frac{s}{z}\right)^3 \bigg( 
          - \frac{1}{3}\H(0,1,1;z)
          + \frac{z}{2}\H(0,1,1;z)
          - \frac{z^2}{6}\H(0,1,1;z)
          - \frac{1}{3}\H(1,1,1;z)\nonumber \\
&& 
          + \frac{z}{2}\H(1,1,1;z)
          - \frac{z^2}{6}\H(1,1,1;z)
          - \frac{z^3}{12}
          + \frac{z^2}{12}
          \bigg) + {\cal O}(s^4)
\nonumber \\
&&
      + \H(0,1,1;z) \G(-z;y)
          + \H(1,1,1;z)\G(-z;y) \; .
\end{eqnarray}
Again, the subsequent expansion of the coefficients in Chebyshev polynomials 
in the rescaled $u$ does yield rational coefficients involving large integer 
numbers in numerator and denominator. The structure of the result is very 
similar to that of the example quoted in the previous subsection. 

To obtain the desired accuracy of 
$3\times 10^{-15}$, one must keep 17 terms in the $s$ expansion of 
$\G(-z,0,1-z,1-z;y)$ and expand the coefficients up to the 10th Chebyshev 
polynomial in the rescaled $u$. Other functions with the same combination 
of indices require up to 18 terms in $s$ and up to the 11th Chebyshev 
polynomial in the rescaled $u$. 

\subsection{The indices $(0,1,1-z,-z)$} 
The 2dHPLs develop their full analytic structure 
if all indices from the set $(0,-z,1-z,1)$ or  $(-z,1-z,1)$  are present 
in $\vec m(z)$: $\G(\vec m(z);y)$ then contains two right cuts in 
$y=1-z$ and $y=1$, as well as one left cut in $y=-z$. 
The irreducible 2dHPLs with this combination of indices are 
chosen to contain always a $(1)$ as rightmost index. Therefore, 
these functions have always the right cut as primary cut, 
with the left cut as secondary cut. 

Again, the left and right cuts are separated according to (\ref{eq:rlsplit}), 
and only $G_+(\vec{a};y)$ needs to be evaluated, since the secondary 
$G_-(\vec{a};y)$ is always expressed in terms of already evaluated functions 
of lower weight. $G_+(\vec{a};y)$ is expanded in $y$ at $y=0$, and $y$ is 
then replaced by the Bernoulli-type variable $s=-\log(1-y/(1-z))$. The 
expansion coefficients take a form similar to the expansion 
coefficients of the 2dHPLs discussed in the two previous subsections. 
The coefficient of $s^n$, which contains terms in $z^{-n}$, can be 
written as an overall factor $z^{-n}$ multiplying a numerator consisting 
of a finite-order polynomial in $z$ plus further $n$-th order polynomials 
in $z$ multiplying HPLs of argument $z$, which contain either indices of the 
subset $(0,1)$ only or indices of the subset $(0,-1)$ only . In $z\to 0$, 
the coefficient and all its derivatives in $z$ are again finite. 
The simultaneous appearance of HPLs of argument $z$ with indices $(0,1)$ 
and of HPLs with indices $(0,-1)$ forbids the introduction of either 
$u=H(1;z)$ or $v=H(-1;z)$ as Bernoulli variable for replacing $z$ in 
the evaluation of the coefficients of the $s$-expansion. 

To evaluate those coefficients, say $c_n(z)$, in an efficient way, we 
write them as $ c_n(z) = c_{n,u}(z) + c_{n,v}(z) $, where $c_{n,u}(z)$ 
corresponds to the $u$-type contributions from HPLs of argument $z$ and 
indices of the subset $(0,1)$ only, and $c_{n,v}(z)$ to the $v$-type 
contributions from HPLs with indices of the subset $(0,-1)$ only. 
The separation of the contributions is evident for the polynomials 
in $z$ multiplying HPLs of either combination of indices. 
The remaining polynomial in $z$, which does not multiply 
any HPL, must also be split into $u$-type and $v$-type contributions, 
which is less evident. The underlying procedure is as follows. 
We recall that the coefficient of $s^n$ is analytic in $z=0$, but has been 
written with an overall factor of $z^{-n}$. The remaining polynomial, as well 
as the terms containing HPLs of $z$, sum up to yield a function 
proportional to $z^n$ for $z\to 0$. Expanding in $z$, around $z=0$, the HPLs 
with indices $(0,1)$, we can identify the terms of the polynomial that 
have to attributed to $c_{n,u}(z)$ to ensure that $c_{n,u}(z)$ is analytic 
in $z=0$. As a result, $c_{n,u}(z)$ is finite for $0\leq z\leq 1$; it is 
however non-analytic in $z=1$, as the HPLs with indices 
$(0,1)$ have a cut at $z=1$. All remaining terms of the polynomial 
are attributed to $c_{n,v}(z)$, which is then analytic in the whole 
interval $0\leq z\leq 1$. 
After performing this separation, $c_{n,v}(z)$ is expanded in terms of 
Chebyshev polynomials in $2z-1$, which yield an accurate description for 
the full interval in $z$. To evaluate $c_{n,u}(z)$, (which is not analytic 
in $z=1$), we split the $z$-interval at $z=1/2$ as in the previous 
subsection. Below this value, $c_{n,u}(z)$ is evaluated by replacing 
$z$ by the Bernoulli-like variable $u=\H(1;z)$, and using the expansion 
in $u$ with subsequent improvement in terms of Chebyshev polynomials. 
Above $z=1/2$, the exact expression is used for the calculation. 

To illustrate the separation of $u$-type and $v$-type terms, we consider 
\begin{eqnarray}
\lefteqn{\G(0,-z,1-z,1;y)}\nonumber \\
&=& 
       + \left(\frac{s}{z}\right)  \biggl(
          - z \H(0,-1;z)
          +  \H(0,-1;z)
          + z\H(0,1;z)
          -  \H(0,1;z) 
          \biggr)
\nonumber \\ &&
       + \left(\frac{s}{z}\right)^2  \left(
          + \frac{z^2}{4} \H(0,-1;z)
          - \frac{1}{4} \H(0,-1;z)
          - \frac{z^2}{4} \H(0,1;z)
          + \frac{1}{4} \H(0,1;z) 
          \right)
\nonumber \\ &&
       + \left(\frac{s}{z}\right)^3  \bigg(
          + \frac{z^2}{18} 
          - \frac{z^3}{9}
          + \frac{z^4}{18} 
          + \frac{1}{9} \H(0,-1;z) 
          - \frac{z}{12} \H(0,-1;z)
          - \frac{z^3}{36} \H(0,-1;z)
\nonumber \\ &&
          - \frac{1}{9} \H(0,1;z) 
          + \frac{z}{12} \H(0,1;z) 
          + \frac{z^3}{36} \H(0,1;z)
          \bigg) 
       + {\cal O} (s^4)\nonumber \\
&&       - \H(0,-1;z)\G(0,-z;y)
          + \H(0,1;z)\G(0,-z;y) \nonumber \\
&=&
       + s  \left(
          - \H(0,-1;z)
          + \frac{1}{z} \H(0,-1;z)
          \right)
       + s  \left(
          + \H(0,1;z)
          - \frac{1}{z} \H(0,1;z) 
          \right)
\nonumber \\ &&
       + s^2  \left( 
         + \frac{1}{4} \H(0,-1;z)
          - \frac{1}{4z^2} \H(0,-1;z)
          + \frac{1}{4z} 
          \right)
       + s^2  \left(
          - \frac{1}{4} \H(0,1;z)
          + \frac{1}{4z^2} \H(0,1;z)
          - \frac{1}{4z} 
          \right)
\nonumber \\ &&
       + s^3  \bigg(
          - \frac{1}{9}
          - \frac{1}{9z^2}
          + \frac{1}{9z} 
          + \frac{z}{18} 
          - \frac{1}{36} \H(0,-1;z)
          + \frac{1}{9z^3} \H(0,-1;z) 
          - \frac{1}{12z^2} \H(0,-1;z)
          \bigg) 
\nonumber \\ &&
       + s^3  \bigg(
          + \frac{1}{36} \H(0,1;z)
          - \frac{1}{9z^3} \H(0,1;z) 
          + \frac{1}{12z^2} \H(0,1;z) 
          + \frac{1}{9z^2}-\frac{1}{18z}
          \bigg) 
       + {\cal O} (s^4) \nonumber \\
&&       - \H(0,-1;z)\G(0,-z;y)
          + \H(0,1;z)\G(0,-z;y) \nonumber \\
&\equiv&
       + s\, c_{1,v}(z) + s\, c_{1,u}(z) 
       + s^2\, c_{2,v}(z) + s^2\, c_{2,u}(z) 
       + s^3\, c_{3,v}(z) + s^3\, c_{3,u}(z)        + {\cal O} (s^4)
\nonumber \\
&&       - \H(0,-1;z)\G(0,-z;y)
          + \H(0,1;z)\G(0,-z;y) \;.
\end{eqnarray} 
An accuracy of $3\times 10^{-15}$ is obtained for $\G(0,-z,1-z,1;y)$ if 
19 terms are kept in the $s$-expansion, the Chebyshev expansion of the 
$v$-type contribution to the coefficients contains 18 terms, and 
the expansion of the $u$-type contribution another 10 terms.
  Other functions with the same combination of indices 
require up to 22 terms in $s$ and up to 19 terms for the 
$v$-type contribution and 11 terms for the $u$-type contribution.

\section{Checks}
\label{sec:checks}
We have carried out several checks of the implementation of the 
algorithm described in the previous section into the {\tt FORTRAN} 
subroutine {\tt tdhpl}. 

An immediate check of the numerical implementation of the 2dHPLs 
is provided by the derivative formula Eq.\ (\ref{eq:derive}). 
Evaluating the left-hand side of Eq.\ (\ref{eq:derive}) numerically,
with a standard symmetric 4-point differentiation formula, and comparing 
it with the right-hand side evaluated directly, we found agreement within 
an accuracy of $10^{-12}$ or better. This accuracy is mainly limited by 
rounding errors induced by the small interval size used in the 
differentiation formula. We used $10^{-4}$ as interval size, which 
implies a theoretical accuracy of about $10^{-16}$. This accuracy is, 
however, reduced to the observed $10^{-12}$ by rounding errors arising 
from taking the difference between function values evaluated 
at interval points spaced by $10^{-4}$ and evaluated only with the 
double precision {\tt FORTRAN} accuracy. 

We also checked the continuity of all 2dHPLs  across the boundaries of 
the two different regions introduced in Section~\ref{sec:numer}, matching 
onto each other at $y=(1-z)/2$. 
Evaluating the 2dHPLs  according to the algorithms appropriate to the 
regions left and right of the boundary, we found both limiting 
values to agree within $3\times 10^{-15}$ or better. Moreover, we evaluated 
the 2dHPLs  in a few points scattered in a small neighbourhood (of size 
$\pm 10^{-10}$) across this boundary, by using for each 
point the two different algorithms 
(used separately in the {\tt FORTRAN} code at each side of the boundaries) 
and then comparing the results. Again, we found agreement within the 
desired accuracy of $3\times 10^{-15}$. 

Up to weight 3, all 2dHPLs can be expressed as Nielsen's generalized 
polylogarithms of suitable argument (see Appendix~\ref{app:nielsen}). 
We have checked that the results produced in the above way 
agree numerically with their expressions in terms of generalized 
polylogarithms, evaluated using {\tt hplog}~\cite{1dhpl} 
in a number of randomly chosen points in $y$ and $z$. 

Finally, some of the two-loop Feynman integrals computed analytically 
in terms of 2dHPLs in~\cite{doublebox} had been computed numerically at 
special points of the 
phase space in~\cite{num}. We find full agreement with these results, 
within the numerical uncertainty quoted in~\cite{num}, which is 
however only 1\%.  This comparison should rather be viewed as 
a verification of the analytical results of~\cite{doublebox} for the
two-loop Feynman integrals.

\section{The subroutine {\tt tdhpl}}
\label{sec:code}
\subsection{Syntax}
The routine {\tt tdhpl} has the following syntax:
\begin{verbatim}
      subroutine tdhpl(y,z,nmax,GYZ1,GYZ2,GYZ3,GYZ4,
     $                          HZ1,HZ2,HZ3,HZ4)
\end{verbatim}

\subsection{Usage}
In calling {\tt hplog}, the user has to supply 
\begin{itemize}
\item[{{\tt y,z}:}] The arguments $y,z$ for which the 2dHPLs  
are to be evaluated.
{\tt y,z} are of type {\tt real*8}. They can take any value inside the 
triangle $0\leq y \leq 1-z$, $0\leq z \leq 1$.
\item[{{\tt nmax}:}] The maximum weight of the 2dHPLs  to be evaluated.
{\tt nw} is of type {\tt integer}. It is limited to $1\leq {\tt nw} \leq 4$. 
\end{itemize}

The output of {\tt tdhpl} is provided in the arrays 
{\tt GYZ1,GYZ2,GYZ3,GYZ4,HZ1,HZ2,HZ3,HZ4}. These have to be 
declared and dimensioned by the user as follows:
\begin{verbatim}
      real*8 HZ1,HZ2,HZ3,HZ4,GYZ1,GYZ2,GYZ3,GYZ4
      dimension HZ1(0:1),HZ2(0:1,0:1),HZ3(0:1,0:1,0:1), 
     $          HZ4(0:1,0:1,0:1,0:1) 
      dimension GYZ1(0:3),GYZ2(0:3,0:3),GYZ3(0:3,0:3,0:3), 
     $          GYZ4(0:3,0:3,0:3,0:3) 
\end{verbatim} 
It should be noted that this declaration is always needed, even if
{\tt tdhpl} is called with {\tt nmax}$<4$. After calling {\tt tdhpl} for 
given arguments {\tt y,z}, the arrays {\tt GYZ1, GYZ2, ...} contain the 
values of the corresponding 2dHPLs  of weight $1,2,\ldots\ $, the arrays 
{\tt HZ1, HZ2, ...} contain the values of the HPLs of argument $z$ and 
indices $(0,1)$, which always appear together with the 2dHPLs of argument 
$y$ and index vector $\vec{m}(z)$ in calculations of Feynman integrals. 
The indices $(0,1,1-z,-z)$, which can appear in the index vector of the 
2dHPLs, 
correspond to the indices $(0,1,2,3)$ of the arrays {\tt GYZ1, GYZ2, ...}. 

The subroutine does not need initialization.

\subsection{Example}
The following example program illustrates how to evaluate 2dHPLs 
up to weight 4 for given values of $y$ and $z$, and to write them out:
\begin{verbatim}
      program test2dhpl
      implicit none
      integer nmax
      integer i,i1,i2,i3,i4
      real*8 y,z
      real*8 HZ1,HZ2,HZ3,HZ4,GYZ1,GYZ2,GYZ3,GYZ4
      dimension HZ1(0:1),HZ2(0:1,0:1),HZ3(0:1,0:1,0:1), 
     $          HZ4(0:1,0:1,0:1,0:1), 
     $          GYZ1(0:3),GYZ2(0:3,0:3),GYZ3(0:3,0:3,0:3), 
     $          GYZ4(0:3,0:3,0:3,0:3)       
      
      nmax = 4

      write (6,*) 'Input y,z:'
      read(5,*) y,z
      call tdhpl(y,z,nmax,GYZ1,GYZ2,GYZ3,GYZ4,HZ1,HZ2,HZ3,HZ4)
      do i1 = 0,3
         write(6,101) i1,GYZ1(i1)
         do i2 = 0,3 
            write(6,102) i1,i2,GYZ2(i1,i2)
            do i3 = 0,3 
               write(6,103) i1,i2,i3,GYZ3(i1,i2,i3)
               do i4 = 0,3 
                  write(6,104) i1,i2,i3,i4,GYZ4(i1,i2,i3,i4)
               enddo
            enddo
         enddo
      enddo
      stop
  101 format('       G(',i2,',y) = ',f18.15)
  102 format('       G(',i2,',',i2,',y) = ',f18.15)
  103 format('       G(',i2,',',i2,',',i2,',y) = ',f18.15)
  104 format('       G(',i2,',',i2,',',i2,',',i2,',y) = ',f18.15)
      end
\end{verbatim}

\section{Numerical examples}
\label{sec:plots}
\begin{figure}[ht]
\begin{center}
\epsfig{file=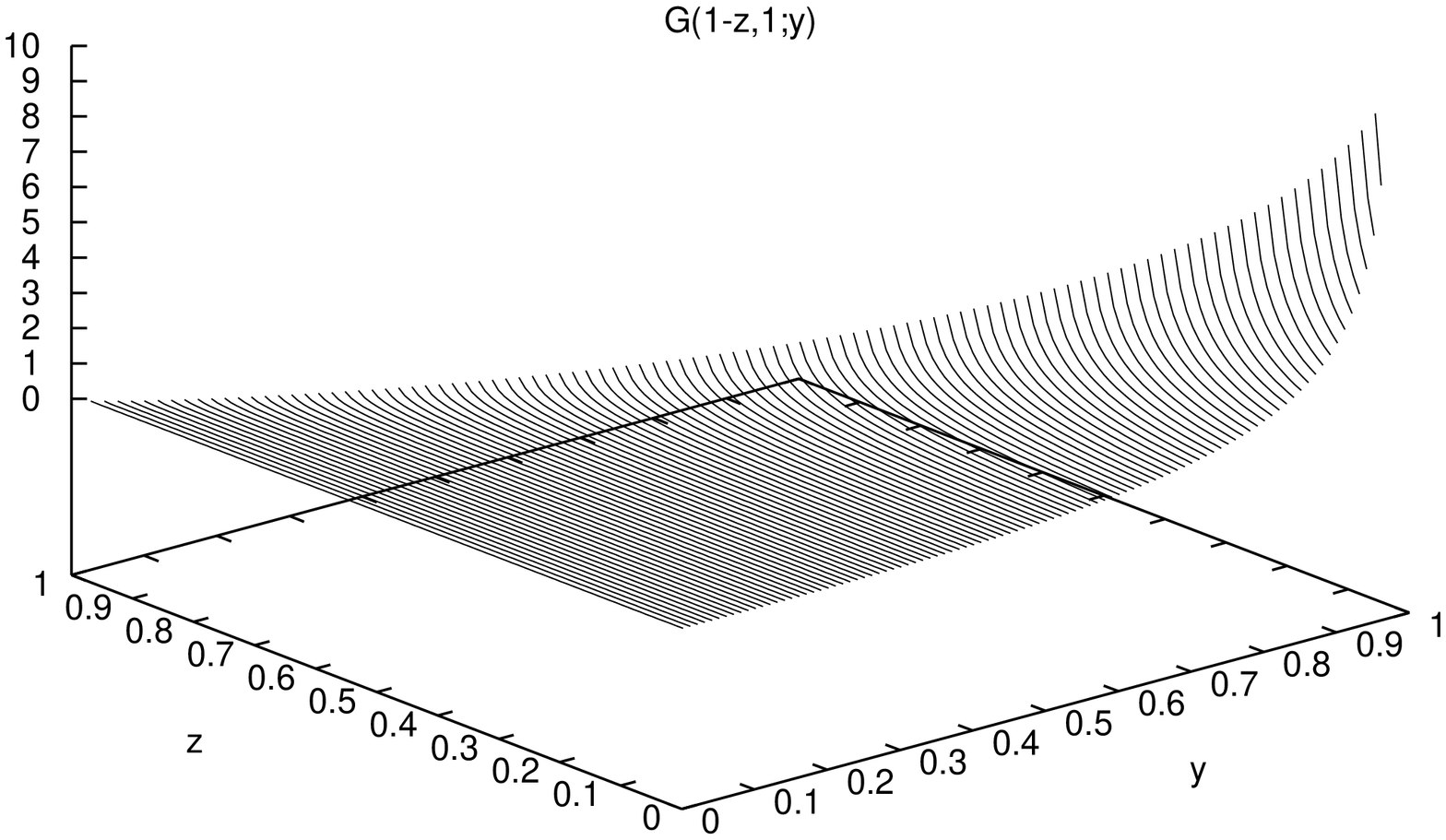,angle=-90,width=9cm}
\vspace{3mm}

\epsfig{file=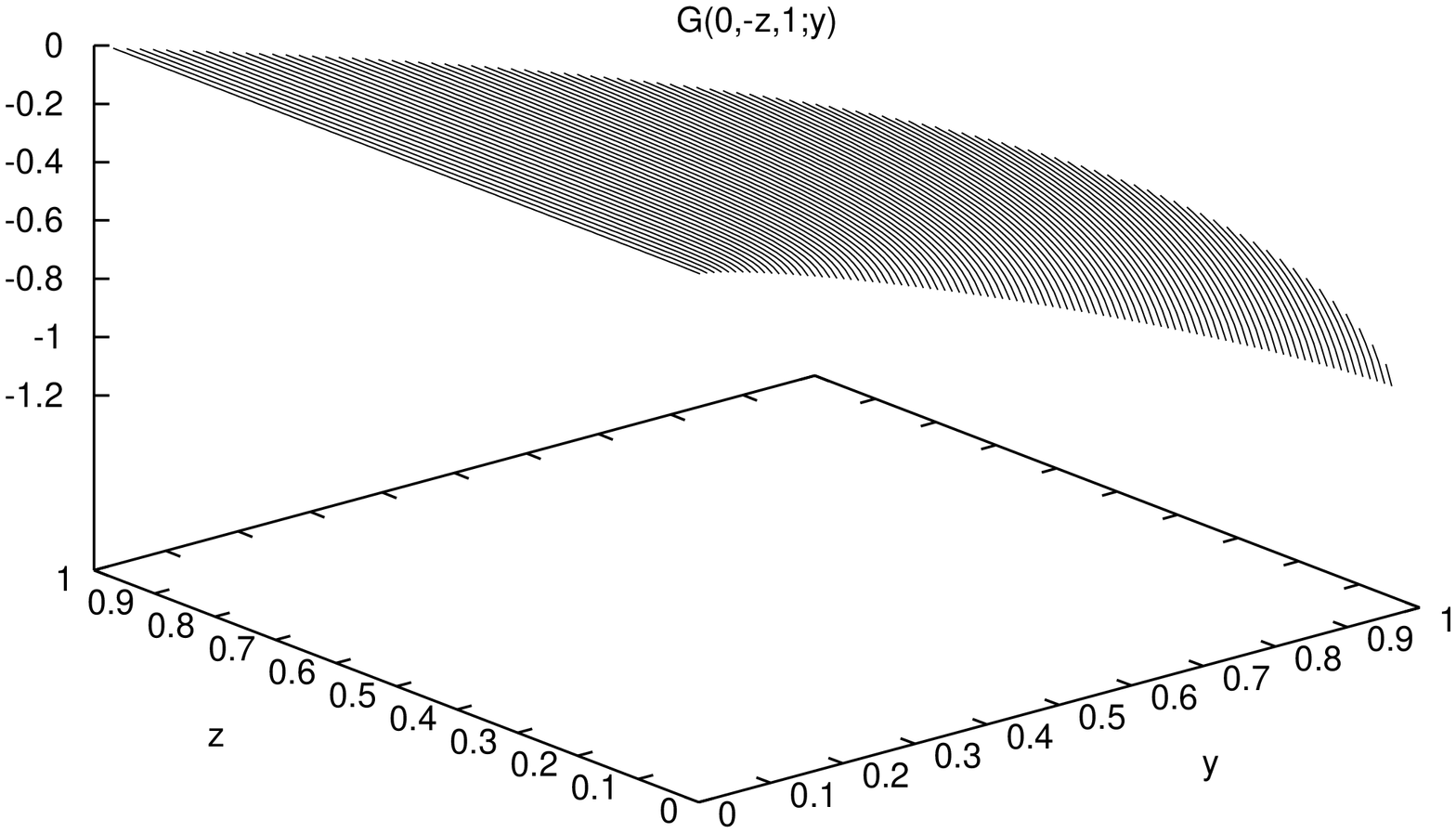,angle=-90,width=9cm}
\vspace{3mm}

\epsfig{file=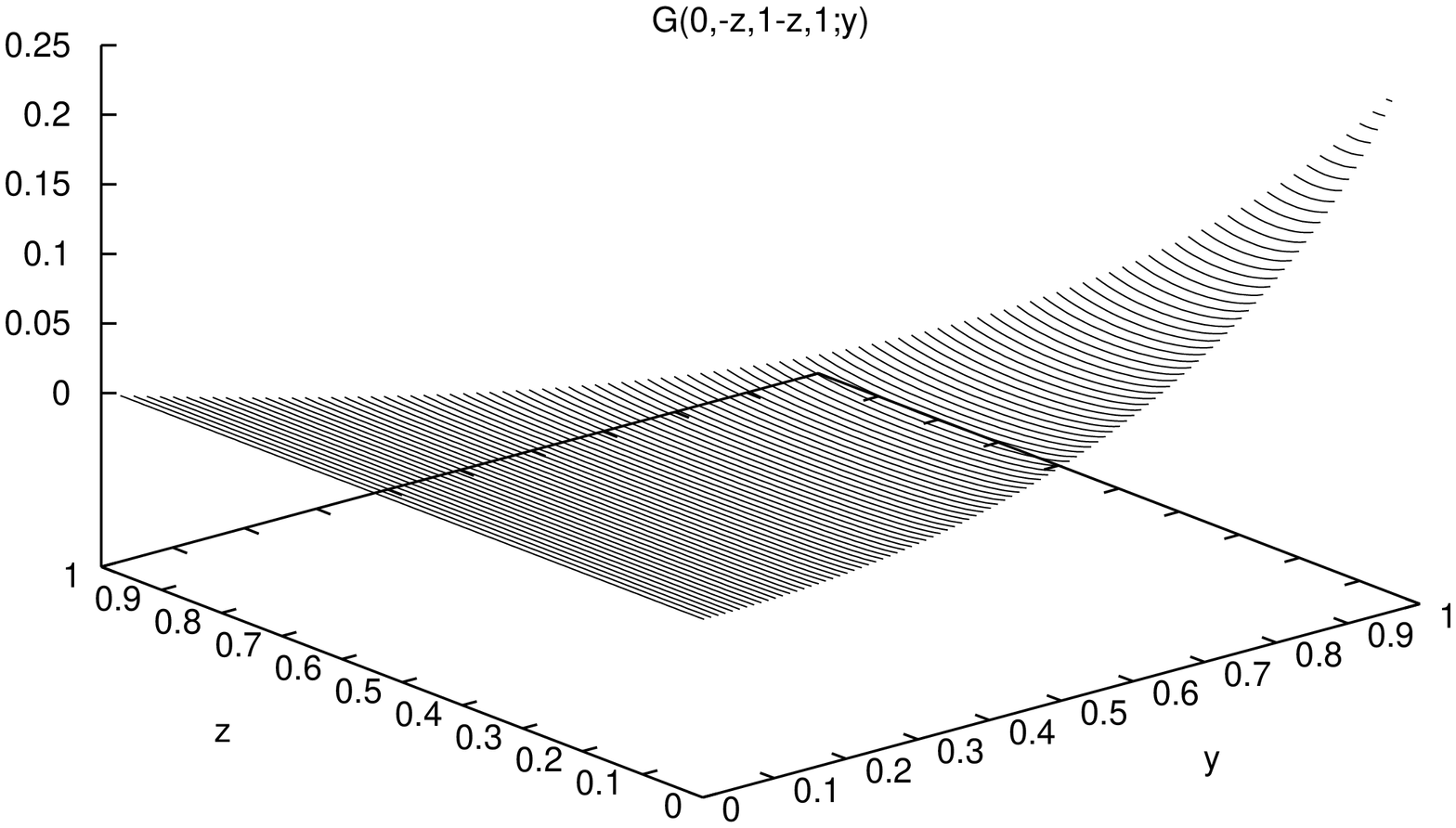,angle=-90,width=9cm}
\end{center}
\caption{Examples for the dependence of 2dHPLs on $y$ and $z$}
\label{fig:plots}
\end{figure}

In Fig.~\ref{fig:plots}, we depict $\G(1-z,1;y)$, $\G(0,-z,1;y)$ and
$\G(0,-z,1-z,1;y)$ as examples of the dependence of the 2dHPLs on $y$ and 
$z$.

\section{Summary} 
In this paper, we have described the routine {\tt tdhpl}, which 
evaluates the two-dimensional harmonic polylogarithms up to weight 4 for 
real arguments in the triangle $0\leq y \leq 1-z$, $\ 0\leq z \leq 1$. 
The evaluation 
is based on a series expansion in terms of appropriately transformed 
expansion parameters for small values of  $y \leq (1-z)/2\;$. The evaluation 
for $y$ in the interval $ (1-z)/2 \leq y \leq 1-z $ 
is based on a transformation formula, relating 
2dHPLs  of arguments $1-y-z$ and $y$. The coefficients 
of the $y$ expansion for small values of $y$, which depend on $z$, are 
then evaluated either in terms of a further power series expansion in $z$ 
or using their exact expression. The convergence of the expansions is 
improved by using Bernoulli-like variables and Chebyshev polynomials. 
The algorithms used and described here can be 
extended to higher weights without further modification, requiring only 
the harmonic polylogarithms up to the desired weight to be known. 

\section*{Acknowledgements}
We are very grateful to Jos Vermaseren for his assistance in the use of 
the algebraic program \FORM~\cite{form}, which was employed intensively 
for generating the code described here. 

We would like to thank Stefan Weinzierl and Johannes Bl\"umlein for 
drawing our attention to related works in the mathematical literature. 

\begin{appendix}
\renewcommand{\theequation}{\mbox{\Alph{section}.\arabic{equation}}}

\section{Comparison of different notations}
\label{app:notation}
\setcounter{equation}{0} 
In this appendix, we compare the different notations used for 
2dHPLs in the literature. In particular, they form a 
special case of the hyperlogarithms of~\cite{Lappo} and the 
multiple polylogarithms of~\cite{gonch}. Since we used a 
different notation for the 2dHPLs in an earlier publication~\cite{doublebox}, 
we also provide the formulae to convert the notation used in~\cite{doublebox}
to the notation employed here. 

A few examples for 2dHPLs written out in the 
different notations are collected in Table~\ref{tab:notate}. 

\subsection{HPLs and previously employed notation for 2dHPLs}
The 2dHPLs were introduced in~\cite{doublebox} as a generalization of the 
HPLs. The one-dimensional HPL $\H(\vec{m}_w;x)$ is described by a 
$w$-dimensional vector $\vec{m}_w$ of parameters and by its argument $x$; 
$w$ is called the weight of $\H$. The HPLs are defined recursively:
\begin{enumerate}
\item
Definition of the  HPLs at $w=1$:
\begin{eqnarray}
\H(1;x) & \equiv & -\ln\, (1-x)\; ,\nonumber \\
\H(0;x) & \equiv & \ln\, x \; ,\nonumber \\
\H(-1;x) & \equiv & \ln\, (1+x) \ ; 
\label{eq:levelone}
\end{eqnarray}
definition of the rational fractions in $x$ 
\begin{eqnarray}
\f(1;x) & \equiv & \frac{1}{1-x} \;, \nonumber \\
\f(0;x) & \equiv & \frac{1}{x} \;, \nonumber \\
\f(-1;x) & \equiv & \frac{1}{1+x} \;,
\end{eqnarray}
such that 
\begin{equation}
\frac{\partial}{\partial x} \H(a;x) = \f(a;x)\qquad \mbox{with}\quad
a=+1,0,-1\;.
\end{equation}
\item For $w>1$:
\begin{eqnarray}
\H(0,\ldots,0;x) & \equiv & \frac{1}{w!} \ln^w x\; ,\\
\H(a,\vec{b};x) & \equiv & \int_0^x \d x^{\prime} 
\f(a;x^{\prime}) \H(\vec{b};x^{\prime})\; , 
\label{eq:inth}
\end{eqnarray}
so that the differentiation formula is, in any case, 
\begin{equation}
\frac{\partial}{\partial x} \H(a,\vec{b};x) = \f(a;x) \H(\vec{b};x)\;.
\label{eq:derivh}
\end{equation}
\end{enumerate}

The notation for 2dHPLs used in~\cite{doublebox} closely resembled the 
above notation of the HPLs by extending the set of fractions by
\begin{eqnarray}
\f(1-z;y) & \equiv & \frac{1}{1-y-z} \;, \nonumber \\
\f(z;y) & \equiv & \frac{1}{y+z} \;, 
\end{eqnarray}
and correspondingly the set of HPLs at weight 1 by
\begin{eqnarray}
\H(1-z;y) &=& - \ln\left(1-\frac{y}{1-z}\right) \ , \nonumber \\
\H(z;y) &=&  \ln\left(\frac{y+z}{z}\right) .
\end{eqnarray}
Allowing $(z,1-z)$ as components of the 
vector $\vec{m}_w$ of parameters, (\ref{eq:inth}) did then define the 
2dHPLs.

However, it turns out that this notation is unpleasant, since the 
rational fractions defined this way are not continuous in $z$ as $z\to 1$: 
indeed one has 
\begin{equation} 
    \lim_{z\to1} \f(1-z;y) = - \f(0;y) \; . 
\end{equation} 
This nuisance is eliminated by the new notation for the 2dHPLs introduced 
in this paper. The relations between the two sets of rational factors are 
\begin{eqnarray}
  \f(1;x) &=& - \g(1;x) \ ,      \nonumber\\ 
  \f(1-z;x) &=& - \g(1-z;x) \ ,  \nonumber\\ 
  \f(z;x) &=& \g(-z;x) \ ,       \nonumber\\  
  \f(0;x) &=& \g(0;z) \ ,
\end{eqnarray} 
and between the 2dHPLs at $w=1$:
\begin{eqnarray}
  \H(1;x) &=& - \G(1;x) \ ,      \nonumber\\ 
  \H(1-z;x) &=& - \G(1-z;x) \ ,  \nonumber\\ 
  \H(z;x) &=& \G(-z;x) \ ,       \nonumber\\  
  \H(0;x) &=& \G(0;z) \ .
\end{eqnarray} 
To convert from the notation of~\cite{doublebox} to the new notation
used here, all $(z)$ in the index vector are to be replaced by $(-z)$, and a 
$(-1)$ is to be multiplied for each occurrence of $(1)$ or $(1-z)$ in the 
index vector. 
\begin{table}[t]
\begin{center}
\begin{tabular}{|c|c|c|c|}\hline
\rule[0mm]{0mm}{5mm}
2dHPL & 2dHPL of \protect\cite{doublebox} & Hyperlogarithm & Multiple 
polylogarithm  
\\[1mm] \hline
\rule[0mm]{0mm}{5mm}
$\G(1-z,-z;y)$ & $-\H(1-z,z;y)$ & $L_0(-z,1-z|y)$ & 
${\rm I}_{1,1}(-z:1-z:y)$ \\[1mm]
$\G(0,0,-z;y)$ & $\H(0,0,z;y)$ & $-$ & 
${\rm I}_{3}(-z:y)$ \\[1mm]
$\G(0,-z,1-z,-z;y)$ & $-\H(0,z,1-z,z;y)$ & $-$ & 
${\rm I}_{1,1,2}(-z:1-z:-z:y)$ \\[1mm]
$\G(1,1,1-z,-z;y)$ & $-\H(1,1,1-z,z;y)$ & $L_0(-z,1-z,1,1|y)$ & 
${\rm I}_{1,1,1,1}(-z:1-z:1:1:y)$ \\[1mm]\hline
\end{tabular}
\caption{Examples of 2dHPLs written in different notations. Some 
2dHPLs cannot be expressed as hyperlogarithms.}
\label{tab:notate}
\end{center}
\end{table}

\subsection{Hyperlogarithms}
The hyperlogarithms of Lappo-Danilevsky~\cite{Lappo} at weight 
$w=1$ are defined by
\begin{equation}
L_b(a_{j_1}|x) \equiv \int_b^x \frac{\d x}{x-a_{j_1}} = \log \frac{x-a_{j_1}}
                                                             {b-a_{j_1}}\;.
\end{equation}
At higher weights, the hyperlogarithms are defined recursively
\begin{equation}
L_b(a_{j_1},a_{j_2},\ldots,a_{j_\nu}|x) 
\equiv \int_b^x \frac{L_b(a_{j_1},a_{j_2},\ldots,a_{j_{\nu-1}}|x)}
      {x-a_{j_\nu}}\, \d x\;.
\end{equation}
Here $b$ is a finite real number, which is different from any of the 
$a_{j_i}$. 

Hyperlogarithms therefore include the subset of the 2dHPLs with all 
indices different from $(0)$:
\begin{equation}
\G(m_1(z),\ldots,m_w(z);y) = L_0(m_w(z),\ldots,m_1(z)|y)\,\qquad 
\mbox{with } m_i(z)\neq 0 \; .
\end{equation}

\subsection{Multiple polylogarithms}
Allowing the lower limit of integration to coincide with one or more 
of the elements of the index vector, hyperlogarithms are 
generalized to multiple polylogarithms. Without loss of generality, 
the lower limit of integration can be taken to be 0. The definition 
of multiple polylogarithms according to Goncharov~\cite{gonch} is then
\begin{eqnarray}
\lefteqn{{\rm I}_{n_1,\ldots,n_m} (a_1:\ldots:a_m:a_{m+1}) \equiv}
\nonumber \\
&&
\int_0^{a_{m+1}} \underbrace{\frac{\d t}{t-a_1}
\circ\frac{\d t}{t}\circ\ldots\circ\frac{\d t}{t}}_{n_1 \mbox{times} }
\circ
\underbrace{\frac{\d t}{t-a_2}
\circ\frac{\d t}{t}\circ\ldots\circ\frac{\d t}{t}}_{n_2 \mbox{times} }
\circ \ldots \circ
\underbrace{\frac{\d t}{t-a_m}
\circ\frac{\d t}{t}\circ\ldots\circ\frac{\d t}{t}}_{n_m \mbox{times} } \; ,
\end{eqnarray}
where
\begin{equation}
\int_0^{x_{m+1}} \frac{\d t}{t-x_1} \circ \ldots \frac{\d t}{t-x_m} \equiv
\int_{0\leq t_1 \leq \ldots \leq t_m \leq t_{m+1}} \frac{\d t_1}{t_1-x_1}
\ldots \frac{\d t_m}{t_m-x_m} \; .
\end{equation}
The subset of the 2dHPLs with trailing index not equal to $(0)$ is 
contained in the multiple polylogarithms:
\begin{equation}
\G(\vec{0}_{n_1},m_1(z),\vec{0}_{n_2},m_2(z),\ldots,\vec{0}_{n_r},m_r(z);y)
= {\rm I}_{(n_r+1),(n_{r-1}+1),\ldots,(n_1+1)}(m_r(z):m_{r-1}(z):\ldots:
m_1(z):y)
\end{equation}
A similar notation has been proposed for the HPLs in~\cite{hpl},
and it is used in the calculations presented in~\cite{moch}. 

\section{Relation to Nielsen's generalized polylogarithms}
\label{app:nielsen}
\setcounter{equation}{0} 
For special values of the indices, it is 
possible to express the 2dHPLs in terms 
of the commonly known Nielsen's generalized polylogarithms~\cite{Nielsen}:
\begin{equation}
{\rm S}_{n,p} (x) = \frac{(-1)^{n+p-1}}{(n-1)! p!} \int_0^1 \d t 
\frac{\log^{n-1} t \log^p (1-xt)}{t}\; ,\qquad  n,p\leq 1 , x\leq 1\; .
\end{equation}
A special case of  ${\rm S}_{n,p} (x)$ is the polylogarithm
\begin{equation}
{\rm Li}_n(x) = {\rm S}_{n-1,1}(x)\; .
\end{equation}
Numerical implementations of the ${\rm S}_{n,p} (x)$ exist in the subroutine
{\tt GPLOG}~\cite{bit} and are widely used.

If the index vector $\vec{m}(z)$ of $\G(\vec{m}(z);y)$ 
contains, besides $(0)$, only one other index, which can appear more than once, 
but only to the right of the rightmost $(0)$, then  $\G(\vec{m}(z);y)$ 
can be expressed as
\begin{eqnarray}
\G(\vec{0}_n,\vec{1}_p;y) &=& (-1)^p {\rm S}_{n,p}(y) \nonumber \\
\G(\vec{0}_n,\stackrel{\longrightarrow}{1-z}_p;y) &=& 
(-1)^p {\rm S}_{n,p}\left(\frac{y}{1-z}\right)  \nonumber \\
\G(\vec{0}_n,\stackrel{\longrightarrow}{-z}_p;y) &=& 
(-1)^p {\rm S}_{n,p}\left(-\frac{y}{z}\right) 
\label{eq:genrel}
\end{eqnarray}

If, besides $(0)$, more than one other index appears in $\vec{m}(z)$, 
$\G(\vec{m}(z);y)$ can be related to generalized 
polylogarithms
only in special cases. 
Relations for all $\G(\vec{m}(z);y)$ exist only up to weight 
$w=3$~\cite{doublebox}. 

The irreducible 2dHPLs at $w=2$ not expressed by (\ref{eq:genrel})
are as follows: 
\begin{eqnarray}
  \G(1-z,1;y) &=& -\frac{1}{2} \ln^2(1-y) + \ln (1-y-z)\ln (1-y)
                 - \Li_2\left(\frac{z}{1-y}\right) 
                 + \Li_2(z) \ , \nonumber \\
  \G(-z,1;y) &=& \ln(1+z) \ln\left(\frac{y+z}{z}\right) 
               + \Li_2\left(\frac{z}{1+z}\right)
               - \Li_2\left(\frac{y+z}{1+z}\right)     \ , \nonumber \\
  \G(z,1-z;y) &=& -\ln(1-z) \ln\left(\frac{y+z}{z}\right)
                 + \Li_2(z) - \Li_2(y+z)   \; .
\end{eqnarray}
Expressions for all irreducible 2dHPLs at $w=3$ in terms of Nielsen's 
generalized polylogarithms are listed in the appendix of~\cite{doublebox}.
\end{appendix}

\end{document}